\documentclass[runningheads,fleqn]{cl2emult}

\usepackage{makeidx}  
\usepackage{graphicx} 
\usepackage{subeqnar} 
\usepackage{multicol} 
\usepackage{cropmark} 
\usepackage{comp}     
\makeindex            

\newtheorem{thm}{Theorem}
\newtheorem{pro}{Proposition}
\newtheorem{fac}{Fact}
\newtheorem{lem}{Lemma}
\newtheorem{cj}{Conjecture}
\newtheorem{cor}{Corollary}
\newtheorem{df}{Definition}

\begin{document}
\title*{The Quantum Entanglement of Binary and Bipolar Sequences}
%
%
%
%
%
\author{Matthew G. Parker\inst{1}
\and V. Rijmen\inst{2}}
%
%
%
\institute{Inst. for Informatikk, University of Bergen, Norway, \\
{\tt matthew@ii.uib.no},
{\tt http://www.ii.uib.no/$\sim$matthew/mattweb.html}
\and Cryptomathic, Lei 8A, B-3000 Leuven, Belgium}

\maketitle              

\begin{abstract}
\index{abstract}
Classification of different forms of quantum \index{entanglement} entanglement is
an active area of research, central to development
of effective \index{quantum computers} quantum computers, and similar to classification
of \index{error-correction codes} error-correction codes, where code
\index{duality} duality is broadened to equivalence under all 'local' \index{unitary transforms}
unitary transforms. We explore
links between entanglement, coding theory, and \index{sequence design} sequence design, by examining
multi-spectra of quantum states under local unitary action,
and show that optimal error-correcting codes and sequences
represent states with high multiparticle entanglement.
\end{abstract}
\section{Introduction}
\label{sec1}
Classification of
multiparticle entanglement of quantum particles is only beginning.
How does one quantify entanglement? There is general agreement about the
entanglement measure for two particles, namely the reduced state \index{entropy} entropy
of the density matrix for the pair \cite{Pop:QE}, but for more than two particles
the measurement criteria are unclear. Moreover, at the moment, a sufficiently refined
measure (or measures) appears to be generally non-computable
(using classical resources) for more than a few particles \cite{Vid:Ent1}.
This paper highlights two partial entanglement measures, namely the
'Linear Entanglement' (LE) (Section \ref{sec7}, Definition \ref{dfXM2}), and
'Stubborness of Entanglement' (SE) (Section \ref{sec8}, Definition \ref{dfXM3}), which is a
sequence of parameters. The paper is aimed at
both coding theorists and sequence designers, and at quantum physicists, and argues that the
best codes and/or sequences can be interpreted as describing multiparticle states
with high entanglement. A binary linear error-correcting code (ECC), ${\bf{C}}$, is often partially described by
its parameters $[n,k,d]$, where $n$ is wordlength, $k$ is code dimension,
and $d$ is minimum \index{Hamming Distance} Hamming Distance \cite{MacW:Cod},
and more generally by its \index{weight hierarchy} weight hierarchy. We show,
by interpreting the length $2^n$ indicator
for ${\bf{C}}$ as an $n$-particle quantum state that, for those states representing
binary linear ECCs, the ECCs with optimal weight hierarchy also have optimal LE
and optimal SE (Theorems \ref{thmXM1} and \ref{thmX4}).
By action of local unitary transform
on the indicator of ${\bf{C}}$, we can also view the quantum state as a bipolar sequence. In this
context a sequence is often partially described by its \index{nonlinear order} nonlinear order, $N$, and
\index{correlation immunity} correlation immunity order, CI (Definitions
\ref{dfx1}, \ref{dfx2}).
We show that $N$ and CI give a lower bound on LE (Theorem \ref{thmX5}).
LE is the $n - \log_2$ of a spectral
'peak' measure of \index{Peak-to-Average Power Ratio} Peak-to-Average Power Ratio (PAR$_l$
(Section \ref{sec7}, Definition \ref{dfXM1})),
which is also an important measure in
telecommunications \cite{Dav:PF,Pat:PF,Par:Gol}. This paper refers both to PAR$_l$ and
to LE, where the two parameters are trivially related (Definition \ref{dfXM2}).
The quantum-mechanical rule of
'local unitary equivalence' is a generalisation of code duality.

We now state the most important results of this paper.
We emphasise quantum states $\vec{s}$ from
the set ${\bf{\ell_p}}$, where ${\bf{\ell_p}}$ is equivalent to the set of binary
linear ECCs.
\begin{itemize}
    \item LE and SE of $\vec{s}$ can be found from the $HI$ multispectra (the multispectra
    resulting from a Local Unitary (LU) transform which is any tensor combination
    of $2 \times 2$ Hadamard and Identity transforms of $\vec{s}$) (Theorems \ref{thmX2}
    and \ref{thmX3}).
    \item Quantum states from ${\bf{\ell_p}}$ which are
    equivalent under LU transform to binary linear ECCs
    with optimum or near-optimum Weight Hierarchy also
   have optimum or near-optimum LE and SE over the set ${\bf{\ell_p}}$
   (Theorem \ref{thmX4}, Corollary \ref{corX2}).
    \item LE$(\vec{s})$ is lower-bounded by an additive combination of the Nonlinear
    Order and Correlation-Immunity of $\vec{s}$ (Theorem \ref{thmX5}).
\end{itemize}
After an initial definition of Entanglement and Measurement (Section \ref{sec2}), we describe binary
and bipolar quantum states (Sections \ref{sec3}, \ref{sec4}, \ref{sec6}), emphasising
states from the set, ${\bf{\ell_p}}$. These are equivalent under LU
transform to binary linear ECCs. We place this analysis
in the context of $HI$ multispectra (Section \ref{sec6}) leading to results
for LE (Section \ref{sec7}) and SE (Section \ref{sec8}). After noting connections
with cryptography (Section \ref{sec8a}), we briefly investigate quadratic states outside
${\bf{\ell_p}}$ (Section \ref{sec9}), and discuss application of the work to measurement-driven
Quantum Computation in Section \ref{sec10}.
\section{Entanglement and Measurement}
\label{sec2}
\subsection{Definitions and Partial Quantification}
A qubit is a two-state particle, $(s_0,s_1)$, meaning
it is in state 0 with complex probability
$s_0$, and state 1 with complex probability $s_1$, such that $|s_0|^2 + |s_1|^2 = 1$.
\begin{df}
Let ${\bf{l_n}}$ be the infinite set of normalised linear vectors which can be written
in the form $(a_0,b_0) \otimes (a_1,b_1) \otimes \ldots \otimes (a_{n-1},b_{n-1})$
\label{dfZ1}
\end{df}
Entanglement exists between two or more particles if their joint probability state cannot be
factorised using the tensor product. More formally,
\begin{df}
Let $\vec{s}$ be an $n$-qubit state. $\vec{s}$ is a pure entangled state if $\vec{s} \not \in
{\bf{l_n}}$. $\vec{s}$ is not entangled if $\vec{s} \in {\bf{l_n}}$.
\label{dfZ2}
\end{df}
{\bf{Example: }} Consider qubits, $x_0$ and $x_1$. Their joint
probability state is given by $ \vec{s} = (s_0,s_1,s_2,s_3) $,
where $s_i$ is complex and $\sum_{i=0}^3 |s_i|^2 = 1$.
If $\vec{s} \in {\bf{l_2}}$,
then $\vec{s}$ is tensor-factorisable and the two qubits are not entangled. Conversely,
if $\vec{s} \not \in {\bf{l_2}}$ then the two qubits are entangled.

If $\vec{s}$ has more than one non-zero entry then the system exists in a 'superposition'
of states. Conventional (classical) computers
only use tensor-factorisable space of physical matter. Humans 'appear'
to only experience this tensor product space, as the exponentially larger entangled
space seems to decohere rapidly. However, perhaps nature also takes advantage
of entanglement and superposition in some way,
as entanglement allows us to manipulate exponentially larger data vectors than possible
classically.
\subsubsection{Partial Quantification of Entanglement}
\begin{itemize}
\item Linear Entanglement: An obvious partial quantification of entanglement is to evaluate
'distance' of a state to the nearest tensor-product state, and Linear Entanglement (LE)
is used in this paper to quantify this distance (Section \ref{sec7}, Definitions
\ref{dfXM1}, \ref{dfXM2}).
\item Stubborness of Entanglement: The second, more refined, partial
entanglement quantification proposed in this paper is the Stubborness of Entanglement
(SE), which is a series of $k'$ parameters, $\beta_j$, specifying entanglement order
after $j$ most-destructive single-qubit measurements on ${\bf{s}}$ (Section
\ref{sec8}, Definition \ref{dfXM3}), where the minimum number of
measurements necessary to completely destroy entanglement is $k'$.
\end{itemize}
\subsubsection{Measurement}
\begin{df}
Let $\vec{s}$ be an $n$-qubit state. Then the vector $\vec{s} = (s_0,s_1,\ldots,s_{2^n-1})$
implies a measurement basis ($\vec{s}$-basis) such that qubit $i$ is measured as $\eta$ with probability
$\sum_{\begin{tiny} \begin{array}{l} k = 0 \\ k_i = \eta \end{array} \end{tiny}}^{2^n-1} |s_k|^2$, where $k$ is
decomposed as $k = \sum_{i=0}^{n-1} k_i2^i$, and $\eta,k_i \in \{0,1\}$. After
measurement, $\vec{s}$ becomes $\vec{s_{x_i = \eta}}$.
\label{dfZ3}
\end{df}
{\bf{Example: }} For $\vec{s} = (s_0,s_1,s_2,s_3)$, there is
a probability $|s_0|^2$,$|s_1|^2$,$|s_2|^2$,$|s_3|^2$ of measuring two qubits in states
$00,01,10,11$, respectively.
As soon as we 'measure' (look at) one or more of the qubits in the $\vec{s}$-basis, then
we destroy the entanglement and superposition of those qubits, effectively projecting
$\vec{s}$ down to a subspace determined by the measurement basis.

For the rest of this paper normalisation of ${\bf{s}}$
is omitted for clarity. Normalisation will always ensure $\sum_{i=0}^{2^n-1} |s_i|^2 = 1$.
\subsection{Entanglement and the Environment}
\begin{figure}
\begin{center}
\includegraphics[width=.5\textwidth]{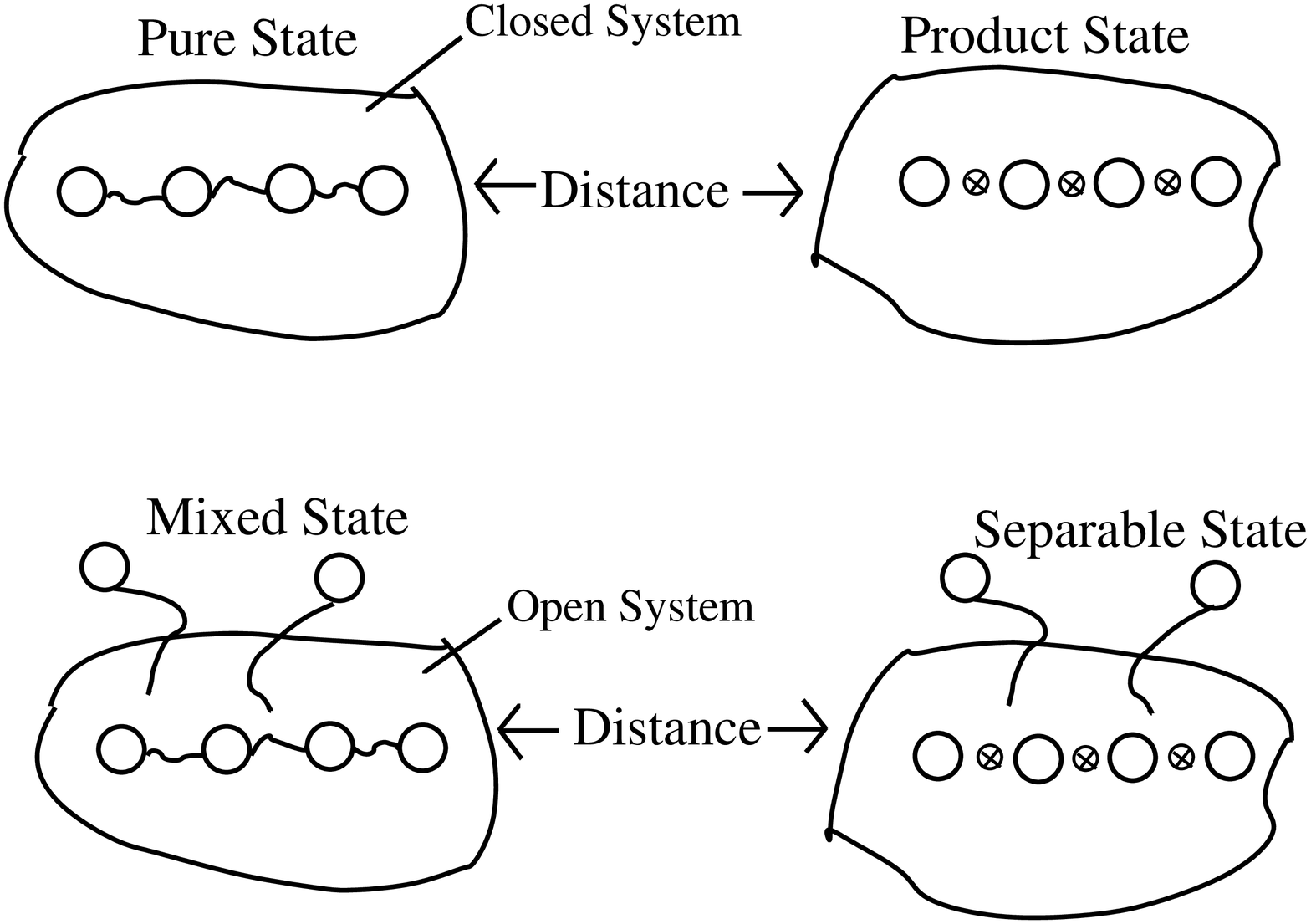}
\caption[]{Pure and Mixed-State Entanglement Measures \label{eps1}}
\end{center}
\end{figure}

Quantum physicists often envisage a more complicated scenario than that described above where one is
trying to measure entanglement of an $n$-qubit system which also has extraneous entanglements with the
environment. It is inaccurate to measure entanglement between $n$ qubits if the system is also
entangled with the environment unless we also take into account the environment. This leads to the
definition of 'mixed states' which are not described by vectors, but by density matrices.
Let $\vec{s}$ be a two qubit pure quantum state $(s_0,s_1,s_2,s_3)$. Then we represent this
state by a $4 \times 4$ density matrix which is the outer product of $\vec{s}$ with $\vec{s^*}$. Thus,
$\vec{\rho_s} = $ \begin{tiny} $\left ( \begin{array}{cccc}
|s_0|^2 & s_0s_1^* & s_0s_2^* & s_0s_3^* \\
s_1s_0^* & |s_1|^2 & s_1s_2^* & s_1s_3^* \\
s_2s_0^* & s_2s_1^* & |s_2|^2 & s_2s_3^* \\
s_3s_0^* & s_3s_1^* & s_2s_3^* & |s_3|^2
\end{array} \right ) $ \end{tiny}.
More generally, an $n$-qubit mixed state is then a statistically
weighted sum of $n$-qubit pure states, and can be described by
$ \vec{\rho} = \sum_i p_i\vec{\rho_{s_i}}$ where $\sum_i p_i = 1$, and $0 \le p_i \le 1$ \cite{Fuc:Qua}.
A local qubit basis is any basis vector from the set ${\bf{l_n}}$
(Definition \ref{dfZ1}).
Although $\vec{\rho}$ may look different under a different local basis,
any entanglement measure for the matrix should be local-basis-independent. The
distance measure of entanglement for mixed states must now measure minimum distance of a mixed state
to the $n$-qubit
tensor product states which may also be entangled with the environment - these are called the
'separable' states. We visualise 'pure' and 'mixed' entanglement distance measures as
in Fig \ref{eps1}.
Mixed-state entanglement is often considered in the literature
\cite{Fuc:Qua,Hor:Ent,Bar:Ent,Wan:Ent}, although definitive entanglement
measures have yet to be agreed upon for more than two particles. For two particles in a mixed
state, Von Neumman entropy of the partial trace, (which is $\mbox{ Tr}(\vec{\rho_r} \log(\vec{\rho_r}))$,
where $\vec{\rho_r}$ is the reduced state density matrix
of $\vec{\rho}$ after 'tracing out' (fixing) one of the two particles) evaluates the distance to the separable states
\cite{Pop:QE,Vid:Ent}. For mixed states of more than two particles, entanglement has been
parameterised using polynomial invariant theory \cite{Bar:Ent,Gra:Ent}, and in many
other ways \cite{Tha:Ent,Pit:Ent,Dur:Ent1,Vir:Ent,Ple:Ent,Eis:Ent,Bras:Ent}, with an
emphasis on asymptotic measures \cite{Hor:Ent,Vid:Ent}. Moreover, some more general approaches to
multiparticle pure state entanglement in an open system are given in
\cite{Tha:Ent,Wan:Ent,Ben:Ent1,Brie:Ent,Eis:Ent}.
A communications engineer could view entanglement of a mixed-state as the 'effectiveness' of a
code in a given channel.
However, in this paper, we only consider pure state entanglement in a closed system -
in other words we assume
little or no entanglement with the environment. Thus we are examining the 'essence' of
entanglement, but side-stepping the more practical issue of its context.
\section{Entanglement and Error-Correction Codes}
\label{sec3}
\begin{pro}
A binary linear $[n,k]$ error-correcting code (ECC), ${\bf{C}}$, can be represented by a length $2^n$
binary indicator, ${\bf{s}}$, where ${\bf{C}} = \{(i_0,i_1,\ldots ,i_{n-1}) | s_i \ne 0
\}$, where $i = \sum_{j=0}^{n-1} i_j2^j$, $i_j \in \{0,1\}$. More specifically,
codeword $(i_0,i_1,\ldots ,i_{n-1})$ in ${\bf{C}}$ occurs with probability $|s_i|^2$.
\label{proZ1}
\end{pro}
In words, the
state vector index is a codeword occurring with probability equal to the magnitude-squared of
the complex value at that index.
A qu{\bf{bit}} state can be re-interpreted as a {\bf{binary}} code, ${\bf{C}}$.
If ${\bf{s}}$ is a binary vector then the state has binary probabilities, and
each codeword is equally likely.
\begin{pro}
The $n$-qubit code represented by $\vec{s}$ satisfies $\vec{s} \not \in \vec{l_n}$
if $\vec{s}$ has error-correction capability (Corollary \ref{corXM1}).
\label{proZ2}
\end{pro}
In words, $\vec{s}$ is entangled and cannot be written as a tensor-product
of length 2 vectors if $\vec{s}$ has error-correction capability.
\subsubsection{Examples}
Consider the two-qubit entangled vector,
$ \vec{s} = (1,0,0,1) $. This state is known as the 'CAT' state in physics literature.
Measurement of the two qubits in the $\vec{s}$-basis produces
states $00$ and $11$ with equal likelihood. $01$ and $10$ are never measured.
$\vec{s}$ is the 'indicator' for the parity-check $[2,1,2]$ code, ${\bf{C}} = \{00,11\}$,
with blocklength $n = 2$, dimension $k = 1$, and Hamming Distance $d = 2$.
By Proposition \ref{proZ2}, $\vec{s}$ is entangled because ${\bf{C}}$ has
error-correction capability.
In contrast, consider the two-qubit unentangled vector,
$ \vec{s} = (1,0,1,0) = (1,0) \otimes (1,1) $.
$\vec{s}$ defines a $[2,1,1]$ code, ${\bf{C}} = \{00,10\}$, which has
minimum Hamming Distance
$1$. This paper also shows that LE and SE are optimised over the set of binary linear
ECCs when ${\bf{C}}$ is optimal (Corollary \ref{corX2}).
Consider the three-qubit entangled vector,
$ \vec{s} = (1,0,0,1,0,1,1,0)  $.
$\vec{s}$ defines a $[3,2,2]$ code, ${\bf{C}} = \{000,011,101,110\}$ which can correct one error in any
of the three bits. It also has optimal LE and SE over the set of length-3 binary linear ECCs.
Of particular interest to physicists are generalised GHZ states which are
indicators over $n$ qubits for binary linear $[n,1,n]$ ECCs. These are
repetition codes, with two codewords, $00\ldots 0$, and $11\ldots 1$.
By Corollary \ref{corXM1}, the high distance, $d = n$,
maximises the SE parameter, $\beta_1$, but the low dimension,
$k = 1$, indicates that only one measurement is necessary to completely destroy entanglement.
\section{Entanglement Equivalence Under Local Unitary Transformation}
\label{sec4}
\begin{figure}
\begin{center}
\includegraphics[width=.4\textwidth]{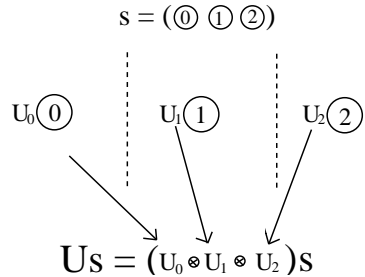}
\caption[]{Local Unitary (LU) Transformation \label{eps2}}
\end{center}
\end{figure}

\begin{figure}
\begin{center}
\includegraphics[width=.5\textwidth]{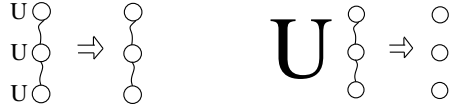}
\caption[]{Changing Entanglement \label{eps3}}
\end{center}
\end{figure}

A unitary transform ${\bf{U}}$ satisfies ${\bf{UU^{\dag} = I}}$ where $\dag$ means
conjugate transpose and ${\bf{I}}$ is identity. Quantum mechanics allows
quantum states to be modified only by pre-multiplication by unitary matrices, or by 'destructive'
measurement (however these destructive measurements could also be notionally described by large unitary matrices
which cover both qubits and environment). By Parseval's theorem, the
unity determinant of a unitary transform matrix preserves energy of
the quantum state but, in general, can make or break entanglement. In this paper, we
are particularly interested in the subclass of unitary matrices, ${\bf{U}}$, which are
entanglement-invariant transformations. Let ${\bf{U}}$ be a $2^n \times 2^n$ unitary transform.
Let $\vec{s}$ be the joint-state vector of $n$ qubits.
\begin{fac}
If ${\bf{U}}$ is tensor decomposable into $2 \times 2$ local unitary (LU) transforms, then
the entanglement of $\vec{s}$ is the same as the entanglement of ${\bf{U}}\vec{s}$.
\label{thm1}
\end{fac}
\begin{df}
If $\exists {\bf{U}}$, where ${\bf{U}}$ is decomposable into $2 \times 2$ local
unitaries, such
that $\vec{s'} = {\bf{Us}}$, then we write $\vec{s'} \equiv_{\mbox{LU}} \vec{s}$
(LU Equivalence).
\label{df1}
\end{df}
Fig \ref{eps2} shows Local Unitary (LU) transformation.
The left-hand (lh) diagram in Fig \ref{eps3} shows invariance of entanglement under LU
transformation, and the right-hand (rh) diagram shows that a tensor-unfactorisable matrix
can break (or make) entanglement.
LU equivalence allows us to develop large equivalence classes of states with
identical entanglement, and view a quantum state from different angles.
If we permit a description of a code such that
codewords can occur with complex probabilities, then we can say,
\begin{pro}
If $\vec{s} \equiv_{\mbox{LU}} \vec{s'}$ then $\vec{s}$ and $\vec{s'}$ represent
'equivalent' codes.
\label{thm1a}
\end{pro}
\begin{pro}
If $\vec{s} \not \equiv_{\mbox{LU}} \vec{s'}$, then
$\vec{s}$ and $\vec{s'}$ represent 'inequivalent' codes.
\label{thm2}
\end{pro}
Code duality is a familiar example of entanglement equivalence. The $2^n \times 2^n$
\index{Walsh-Hadamard Transform} Walsh-Hadamard Transform (WHT) is
\begin{tiny} $\bigotimes_{i=0}^{n-1} \left ( \begin{array}{cc}
1 & 1 \\
1 & -1
\end{array}\right )$ \end{tiny}:
\begin{pro}
Let ${\bf{C}}$ and ${\bf{C^{\perp}}}$ be binary linear ECCs
described by the indicators $\vec{s}$
and $\vec{s'}$, respectively. Then $\vec{s} \equiv_{\mbox{LU}} \vec{s'}$,
where the LU transform is the WHT.
\label{thm3}
\end{pro}
{\bf{Example: }}
The $[3,1,3]$ binary linear ECC, $\vec{s'} = (1,0,0,0,0,0,0,1)$, is obtained from
the $[3,2,2]$ binary linear ECC,
$\vec{s} = (1,0,0,1,0,1,1,0)$, by application of the $8 \times 8$ WHT
\cite{MacW:Cod,Roy:Dual}.
\section{$HI$ Multispectra and the Set, ${\bf{\ell_p}}$}
\label{sec6}
\subsection{Theory}
Entanglement of quantum states is invariant under {\underline{any}} LU transform, but this paper
emphasises the spectra of a subset of quantum states after transform by
LU tensor products of ${\bf{H}}$ and ${\bf{I}}$,
where
${\bf{H}} = $ \begin{tiny} $\left ( \begin{array}{cc}
1 & 1 \\
1 & -1
\end{array}\right )$ \end{tiny}, and
${\bf{I}} = $ \begin{tiny} $\left ( \begin{array}{cc}
1 & 0 \\
0 & 1
\end{array}\right )$ \end{tiny}.
We call this set of transforms the '$HI$ Transform', and the associated spectra the
'$HI$ multispectra'. In this section we identify a set of states, ${\bf{\ell_p}}$,
which are LU equivalent to the set of binary linear ECCs, via $HI$ transforms
(Theorem \ref{thma6}). In later sections we show that, for
states from ${\bf{\ell_p}}$, the LE and SE
can be found from the $HI$ multispectra (Theorems \ref{thmX2},\ref{thmQ},\ref{thmX3},
\ref{thmX4}).
To describe these spectra we require the Algebraic Normal Form
(ANF) for an associated function \cite{MacW:Cod}. For instance, $ \vec{s} = (1,0,0,1,0,1,1,0)  $ can alternatively
be described by the boolean function, $s(x_0,x_1,x_2) = x_0 + x_1 + x_2 + 1$, where
$s(x_0,x_1,x_2)$ can be interpreted as a sequence by concatenating the
function evaluations at $x_2x_1x_0 = $ $000$, $001$, $010$, $\ldots$, $111$ (a lexicographic ordering)
to form the sequence $10010110$. We further propose Algebraic Polar Form (APF) which
separates magnitude (binary), and phase (bipolar) properties of $\vec{s}$,
$$ \vec{s} = s({\bf{x}}) = m({\bf{x}})(-1)^{p({\bf{x}})} $$
where $m({\bf{x}})$ and $p({\bf{x}})$ are both boolean Algebraic Normal Forms (ANFs) in
$n$ binary variables ${\bf{x}} = \{x_0,x_1,x_2,\ldots,x_{n-1}\}$. The
coefficients of $s({\bf{x}})$ are in the set $\{-1,0,1\}$ (we consider more general alphabets in
future papers). Normalisation is ignored
so that it is always assumed that the magnitude-squareds of the coefficients of $s({\bf{x}})$ sum to
one. For example:
\begin{small} $$ \begin{array}{c}
s({\bf{x}}) = (x_0 + x_1 + x_2 + 1)(x_1 + x_3)(-1)^{(x_1x_2 + x_0 + 1)} = \\
0,0,0,1,0,0,1,0,-1,0,0,0,0,1,0,0
\end{array} $$ \end{small}
where '$+$' and '$-$' are mod 2 operations.
$m({\bf{x}})$ is decomposed as $ m({\bf{x}}) = \prod_k h({\bf{x}})_k $.
For these
preliminary investigations, we are interested in $m({\bf{x}})$
comprising $h({\bf{x}})_k$ of degree 1, and $p({\bf{x}})$ of degree $\le 2$.
APFs are useful because, for graphs with many qubits but relatively few ${\overline{\mbox{XOR}}}$ or AND
connections (low-density)
we do not wish to operate explicitly on a very large state vector. The APF allows us to implicitly
act on this vector via the compact APF description
\footnote{LDPC and Turbo-Decoding strategies also exploit 'low-density'
\cite{Par:QFG,Ksc:Fac,Berr:Tur}, and it
is hoped that APF in conjunction with graph-based models will be useful for the
construction and analysis of many-qubit entangled systems \cite{Sch:QG,Dur:Ent2,Woo:Ent}.}.
Only main results are presented here. Appendix \ref{App} provides proofs and further
subsidiary theorems.
\begin{df}
"$H$ acting on $i$" means the action of the transform,
${\bf{I}} \otimes \ldots \otimes {\bf{I}} \otimes {\bf{H}} \otimes {\bf{I}} \otimes \ldots \otimes {\bf{I}}$ on $\vec{s}$,
where ${\bf{H}}$ is preceded by $i$ ${\bf{I}}$ matrices, and followed by $n - i - 1$ ${\bf{I}}$ matrices.
We write this as $H(i)$, or $H(i)[\vec{s}]$.
\label{dfa3}
\end{df}
$\prod_{i \in {\bf{T}}} H(i)$ is the action of $H$ on the subset of qubits $i$ represented by the integers
in ${\bf{T}}$. This action can happen in any order as each $H(i)$ acts {\bf{locally}} only on qubit
$i$.
\begin{df}
Let $s({\bf{x}}) = m({\bf{x}})(-1)^{p({\bf{x}})}$ be a binary APF, where
$\deg(p({\bf{x}})) \le 2$, and $m({\bf{x}})$ is such that $\deg(h({\bf{x}})_k)  = 1$, $\forall k$.
We refer to such an $s({\bf{x}})$ as a 'binary spectra APF'.
Let ${\bf{\Theta}}$ be the set of all binary spectra APF.
\label{dfa4}
\end{df}
\begin{thm}
${\bf{\Theta}}$ is closed under the action of $\prod_{i \in {\bf{T}}} H(i)$, $\forall$
${\bf{T}}$, where ${\bf{T}} \subset \{0,1,\ldots,n-1\}$.
\label{thma4}
\end{thm}
\begin{proof} Section \ref{App}.
 \hspace{5mm} \mbox{ }\rule{2mm}{3mm} \end{proof}
Theorem \ref{thma4} implies that the $HI$ multispectra (including WHT) of a binary spectra APF
is at most three-valued.
One subset of ${\bf{\Theta}}$ is $ s({\bf{x}}) = (-1)^{p({\bf{x}})}$.
Another subset of ${\bf{\Theta}}$ is $ s({\bf{x}}) = m({\bf{x}})$,
where $m({\bf{x}})$ is a product of linear functions.
\begin{df}
Let ${\bf{T_C}}$, ${\bf{T_{C^{\perp}}}}$ be integer sets chosen so that
${\bf{T_C}} \cap {\bf{T_{C^{\perp}}}} = \emptyset$, and
${\bf{T_C}} \cup {\bf{T_{C^{\perp}}}} = \{0,1,\ldots,n-1\}$. This is a bipartite
splitting of $\{0,1,\ldots,n-1\}$. Let us also partition the variable set ${\bf{x}}$ as
$ {\bf{x}} = {\bf{x_C}} \cup {\bf{x_{C^{\perp}}}} $,
where ${\bf{x_C}} = \{x_i | i \in {\bf{T_C}} \}$,
and ${\bf{x_{C^{\perp}}}} = \{x_i | i \in {\bf{T_{C^{\perp}}}} \}$.
\label{dfa6}
\end{df}
\begin{df}
${\bf{\ell_p}}$ is the subset of all $s({\bf{x}})$ from ${\bf{\Theta}}$
of the form
$s({\bf{x}}) = (-1)^{p({\bf{x}})}$, where
$ p({\bf{x}}) = \sum_k q_k({\bf{x_C}})r_k({\bf{x_{C^{\perp}}}}) $,
where $\deg(q_k({\bf{x_C}})) = \deg(r_k({\bf{x_{C^{\perp}}}})) = 1$ $\forall k$,
and where $x_i \in p({\bf{x}})$,
$\forall$ $i \in \{0,1,\ldots,n-1\}$. We refer to
${\bf{\ell_p}}$ as the set of 'bipartite quadratic bipolar' states.
\label{dfa7}
\end{df}
\begin{thm}
If $\vec{s} \in {\bf{\ell_p}}$, then the action of $\prod_{i \in {\bf{T}}} H(i)$
on $\vec{s}$ gives $s'({\bf{x}}) = m({\bf{x}})$, for ${\bf{T}} = {\bf{T_C}}$, or
${\bf{T}} = {\bf{T_{C^{\perp}}}}$. $\vec{s'}$ is the binary indicator
for a binary linear $[n,n-|{\bf{T}}|,d]$ error correcting code, ${\bf{C}}$, (or ${\bf{C^{\perp}}}$),
if ${\bf{T_C}}$, (or ${\bf{T_{C^{\perp}}}}$) is used.
\label{thma6}
\end{thm}
\begin{proof}
Section \ref{App}.
 \hspace{5mm} \mbox{ }\rule{2mm}{3mm} \end{proof}
\begin{small}
{\bf{Example: }} Let ${\bf{T_C}} = \{0,2,5,7\}$ and ${\bf{T_{C^{\perp}}}} = \{1,3,4,6\}$.
Let, \newline
$s({\bf{x}}) = (-1)^{x_0x_1 + x_0x_3 + x_0x_4 + x_1x_2 + x_1x_5 + x_2x_3 + x_2x_6 + x_3x_7 + x_4x_5
 + x_4x_7 + x_5x_6 + x_6x_7}$. Then, for ${\bf{T}} = {\bf{T_C}}$,
$s'({\bf{x}}) = (x_0 + x_1 + x_3 + x_4 + 1)(x_1 + x_2 + x_3 + x_6 + 1)(x_1 + x_4 + x_5 + x_6 + 1)
(x_3 + x_4 + x_6 + x_7 + 1)$ is an indicator for the $[8,4,4]$ binary linear code.
\end{small} \newline
We now consider certain $p({\bf{x}})$ of any degree.
\begin{df}
${\bf{\aleph}}$ is the set of $s({\bf{x}})$ of the form
$s({\bf{x}}) = (-1)^{p({\bf{x}})}$, where
$ p({\bf{x}}) = \sum_k q_k({\bf{x_C}})r_k({\bf{x_{C^{\perp}}}}) $,
where $\deg(q_k({\bf{x_C}})) = 1$ $\forall k$, $r_k({\bf{x_{C^{\perp}}}})$
is of arbitrary degree $\forall k$, and where $x_i \in p({\bf{x}})$,
$\forall$ $i \in \{0,1,\ldots,n-1\}$.
\label{dfa8}
\end{df}
\begin{cj}
If $\vec{s} \in {\bf{\aleph}}$ then the action
of $\prod_{i \in {\bf{T_C}}} H(i)$ on $\vec{s}$ gives 
$s'({\bf{x}}) = m'({\bf{x}})$, where $m'({\bf{x}})$ is a binary polynomial.
We can interpret $\vec{s'}$ as a
binary error correcting code (linear or nonlinear, as appropriate).
\label{cj1}
\end{cj}
\subsection{Examples}
\label{sec5}
We now discuss a few examples of the LU equivalence via $HI$ transforms between quadratic
bipartite bipolar sequences (the set ${\bf{\ell_p}}$) and the binary
indicators for binary linear ECCs, highlighting graphical representations. \newline
{\bf{Example 1: }} Let
$\vec{s} = $ \begin{tiny} $(+++-++-++++---+-)$ \end{tiny},
where '$+$' $ = 1$, and '$-$' $ = -1$. Then $\vec{s} = (-1)^{0001001000011101}
\in {\bf{\ell_p}}$,
In ANF this is,
$s(x_0,x_1,x_2,x_3) = (-1)^{x_0x_1 + x_1x_2 + x_2x_3}$. This quadratic ANF
satisfies a bipartite splitting where ${\bf{T_C}} = \{x_0,x_2\}$ and
${\bf{T_{C^\perp}}} = \{x_1,x_3\}$. We can apply $H(0)H(2)$
to $\vec{s}$ to get $\vec{s'} = (1000000100011000) = (x_0 + x_1 + 1)(x_1 + x_2 + x_3 +
1)$, which is the binary indicator for a $[4,2,2]$ binary linear block code, ${\bf{C}}$.
Alternatively, we can apply $H(1)H(3)$
to $\vec{s}$ to get $\vec{s''} = (1001000000000110) = (x_0 + x_1 + x_2 + 1)(x_2 + x_3 +
1)$, which is the binary indicator for the $[4,2,2]$ binary linear block code,
${\bf{C^{\perp}}}$.

\begin{figure}
\begin{center}
\includegraphics[width=.7\textwidth]{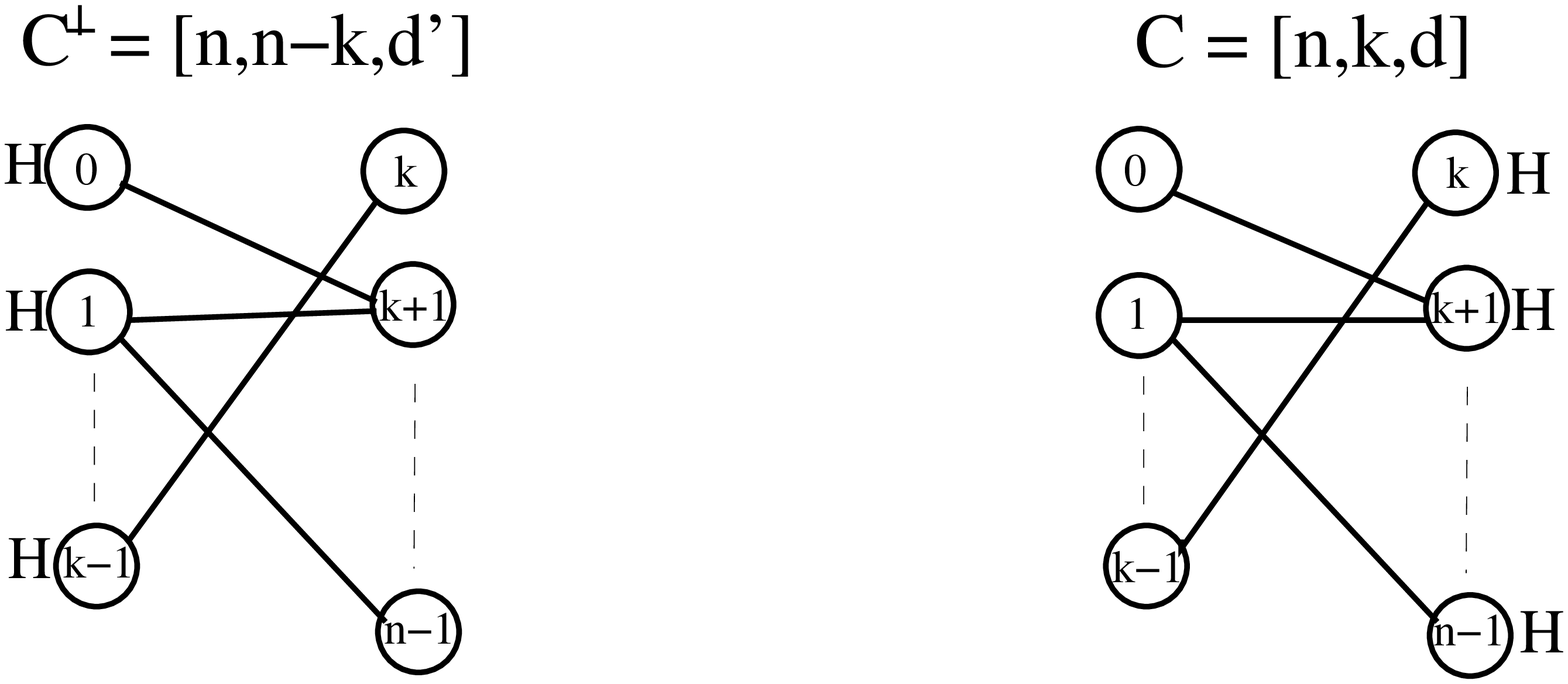}
\caption[]{Quadratic Bipolar Bipartite State from ${\bf{\ell_p}}$ Before/After
Applying ${\bf{H}}$ Operators \label{eps4}}
\end{center}
\end{figure}

\begin{figure}
\begin{center}
\includegraphics[width=.6\textwidth]{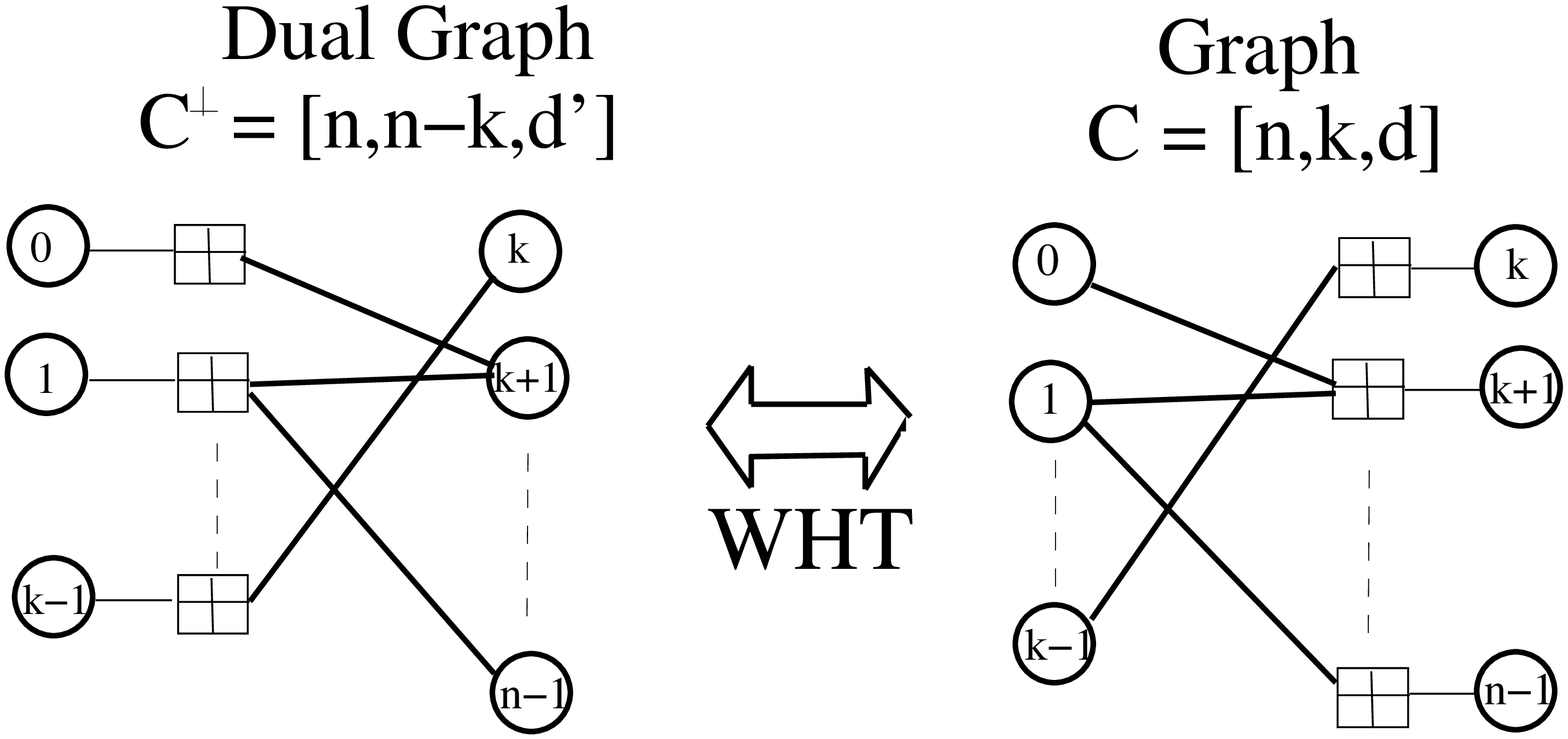}
\caption[]{Binary Indicator State (Factor Graph) \label{eps5}}
\end{center}
\end{figure}

\begin{figure}
\begin{center}
\includegraphics[width=1\textwidth]{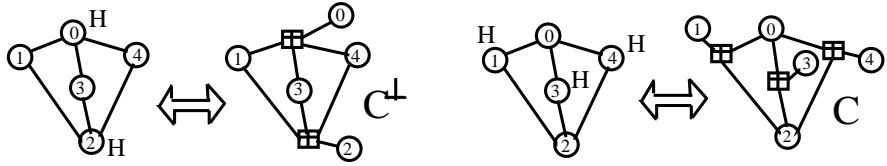}
\caption[]{Bipolar to Binary Equivalence (Example 2) \label{eps6}}
\end{center}
\end{figure}

We can represent members of ${\bf{\ell_p}}$
graphically, as in Fig \ref{eps4}, where a circle is a qubit, and a line implies the
existence of a quadratic term comprising the qubits at either end of the line.
Applying ${\bf{H}}$
to all qubits on the left, or to all qubits on the right, converts the bipolar sequence
to the binary indicator representing ${\bf{C}}$, or
${\bf{C^{\perp}}}$, respectively. The number of ${\bf{H}}$ operators applied
determines the number of parity bits for the code (Theorem \ref{thma6}).
Entanglement of a member of ${\bf{\ell_p}}$ is identical
to the LU-equivalent indicators for ${\bf{C}}$ and
${\bf{C^{\perp}}}$
\footnote{A potential source of confusion is that a cryptographer often equates binary sequence
$abcde$ with bipolar sequence $(-1)^{abcde}$ via the operation $2\{0,1\}-1 = \{-1,1\}$. This
non-unitary equivalence is {\bf{forbidden}} in quantum systems.}.
After appropriate Walsh-Hadamard rotations to the bipartite bipolar sequence, the
resultant binary indicator can also be represented graphically, as shown in
Fig \ref{eps5}, where squares with crosses represent ${\overline{\mbox{XOR}}}$.
This '\index{Factor Graph} Factor Graph' representation is currently the subject of much research in the
context of \index{Low Density Parity Check} Low Density Parity Check (LDPC) codes for
iterative decoding \cite{Par:QFG,Ksc:Fac}, and the
Factor Graph form that arises from selective Walsh-Hadamard rotations of a
quadratic bipartite bipolar
sequence is the '\index{Normal Realisation} Normal Realisation', as recently described in \cite{For:GN}.
\newline
{\bf{Example 2: }}
The sequence \begin{tiny}$+++-++-++-++-++++-++-++++++-++-+$ \end{tiny}
has ANF
$s(x_0,x_1,x_2,x_3,x_4) = (-1)^{x_0x_1 + x_0x_3 + x_0x_4 + x_1x_2 + x_2x_3 + x_2x_4}$ and is equivalent via
action of $H(1)H(3)H(4)$ to the indicator,
$s'(x_0,x_1,x_2,x_3,x_4) = (x_0 + x_1 + x_2 + 1)(x_0 + x_2 + x_3 + 1)(x_0 + x_2 + x_4 + 1)$, for a
$[5,2,2]$ binary linear ECC, ${\bf{C}}$.
Alternatively $s(x_0,x_1,x_2,x_3,x_4)$ is equivalent via action of $H(0)H(2)$
to the indicator, $s''(x_0,x_1,x_2,x_3,x_4) = (x_0 + x_1 + x_3 + x_4 + 1)(x_1 + x_2 + x_3 + x_4 +
1)$, for a $[5,3,2]$ binary linear ECC, ${\bf{C^{\perp}}}$.
We illustrate Example 2 in Fig \ref{eps6} where the lh-side is a bipolar graph, and the rh-side
is a binary graph for the associated binary code.

\begin{figure}
\begin{center}
\includegraphics[width=.3\textwidth]{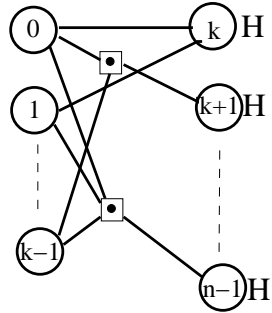}
\caption[]{Bipolar Bipartite Sequences of any ANF Degree \label{eps7}}
\end{center}
\end{figure}

The LU equivalence between the set ${\bf{\ell_p}}$ and binary linear ECCs
also extends to LU equivalence between {\bf{higher-degree}} bipartite bipolar sequences
and binary {\bf{nonlinear}} ECCs. A bipolar bipartite sequence of any ANF degree is
shown in Fig \ref{eps7}, where the black dot in a square represents AND,
and the sequence is LU equivalent to a binary indicator
for a binary ECC (linear if the bipolar degree $\le 2$, nonlinear otherwise), by
application of ${\bf{H}}$ to all qubits on the rh-side (not the lh-side).
The dimension of the associated binary code, $k$, is then given by the number of
qubits on the lh-side. The general rule is that ${\bf{H}}$ is applied to one,
and only one, variable in each product term of the bipolar ANF (Conjecture \ref{cj1}), and in Fig \ref{eps7}
no two qubits on the right occur in the same product term.
Fig \ref{eps7} shows, as an example, the ANF,
$(-1)^{x_0x_1x_{k-1}x_{n-1} + x_0x_{k-1}x_{k+1} + x_1x_k + x_0x_k + p'({\bf{x}})}$, for some bipartite
$p'({\bf{x}})$. \newline
{\bf{Example 3: }}
The nonlinear $[16,8,6]$ Nordstrom-Robinson binary ECC is LU-equivalent to
a bipolar bipartite sequence (a member of ${\bf{\aleph}}$, Conjecture
\ref{cj1}) with an ANF which comprises 96 cubic terms and 40 quadratic terms,
and where $|T_C| = |T_{C^{\perp}}| = 8$.
The binary linear $[16,8,4]$ ECC, described by the
40 quadratic terms can be 'doped' with 96 cubic terms to increase Hamming Distance
from 4 to 6.
We have the following nonlinear subcodes with underlying linear subcodes, which are
LU-equivalent under a combination of ${\bf{H}}$ and ${\bf{I}}$ operators to
bipolar sequences with cubic $+$ quadratic, or quadratic ANFs, respectively.
\begin{small}
$$ \begin{array}{lllcl}
\mbox{Nonlinear} & [16,8,6] & \Leftarrow HI \Rightarrow & \mbox{96 cubics} & \mbox{40 quads} \\
\mbox{Linear} & [16,8,4] & \Leftarrow HI \Rightarrow & - & \mbox{40 quads} \\
\mbox{Subcode:} & & & & \\
\mbox{Nonlinear} & [12,4,6] & \Leftarrow HI \Rightarrow & \mbox{16 cubics} & \mbox{20 quads} \\
\mbox{Linear} & [12,4,6] & \Leftarrow HI \Rightarrow & - & \mbox{20 quads} \\
\mbox{Subcode:} & & & & \\
\mbox{Linear} & [10,2,6] & \Leftarrow HI \Rightarrow & - & \mbox{10 quads}
\end{array} $$
\end{small}
\section{PAR$_l$ and Linear Entanglement (LE)}
\label{sec7}
\subsection{Definitions}
Partial quantification of entanglement for $\vec{s}$ of length $2^n$ ($n$ qubits) is
achieved by measuring maximum possible correlation of $\vec{s}$ with {\bf{any}} length
$2^n$ 'linear' sequence, $\vec{l_n}$. This correlation
maximum can be expressed as a 'Peak-to-Average-Power-Ratio' (PAR$_l$),
\begin{df}
$$ \mbox{PAR$_l$}(\vec{s}) = 2^n \mbox{max}_{\vec{l}}(|\vec{s \cdot l}|^2) $$
where $\vec{l}$ is any normalised linear sequence from the set,
$\vec{l_n} = \{(a_0,b_0) \otimes (a_1,b_1) \otimes \ldots \otimes (a_{n-1},b_{n-1})\}$, and
$\cdot$ means 'inner product' \cite{Par:Gol}.
\label{dfXM1}
\end{df}
Linear Entanglement (LE) is then defined as,
\begin{df}
$$ \mbox{LE}(\vec{s}) = n - \log_2(\mbox{PAR$_l$}(\vec{s})) $$
\label{dfXM2}
\end{df}
$\vec{s}$ is completely uncorrelated with all
linear sequences and wholly correlated with a particular linear sequence (unentangled)
when its PAR$_l$ is $1$ and $2^n$, respectively. PAR$_l$ is LU-invariant, and is a
natural generalisation of well-known spectral measures, as it includes WHT spectra and,
more generally, all
one and multi-dimensional complex Discrete Fourier spectra, as subspectra.
We can alternatively express PAR$_l$ as,
\begin{df}
\begin{small}
Let $\vec{s'} = (s_0',s_1',\ldots,s_{2^n-1}') =
(\bigotimes_{j \in {\bf{T}}} {\bf{U}}(j))\vec{s}$, where
\begin{small} $ {\bf{U}}(j) =
\left ( \begin{array}{cc}
\cos \theta_j & \sin \theta_j e^{iw_j} \\
\sin \theta_j e^{-iw_j} & -\cos \theta_j
\end{array} \right ) $ \end{small}, $i^2 = -1$, and $\theta_j$ and $w_j$ can take any real values, $\forall
j$. Then,
$$ \begin{array}{c}
\mbox{PAR$_l$}(\vec{s}) = 2^n \mbox{max}_t(|s_t'|^2),
\hspace{5mm} \mbox{ }\forall {\bf{T}} \subset \{0,1,\ldots,n-1\} \end{array} $$ \end{small}
\label{dfZ5}
\end{df}
We have written software to compute PAR$_l$ for any $\vec{s}$. However the
computation is too large for more than about 5 or 6 qubits.
\subsection{PAR$_l$ for States from ${\bf{\ell_p}}$}
\begin{df}
$$ \mbox{PAR}(\vec{s}) = 2^n \mbox{max}_i(|s_i|^2) $$
\label{dfZ6}
\end{df}
\begin{thm}
If $\vec{s} \in {\bf{\ell_p}}$, then $\vec{s}$
is LU equivalent to the indicator for an $[n,k,d]$ binary linear code, and,
$$ \mbox{PAR$_l$}(\vec{s}) \ge 2^{r}, \hspace{3mm} \mbox{ where } r = \mbox{max}(k,n-k) $$
\label{thmx1}
\end{thm}
\begin{proof}
Without loss of generality let $k \le n-k$. The indicator for the $[n,k,d]$
code has $2^k$ equal magnitude non-zero coefficients. By Parseval's Theorem and normalisation, the PAR of
such an indicator must be $2^{n-k}$.
 \hspace{5mm} \mbox{ }\rule{2mm}{3mm} \end{proof}
Theorem \ref{thmx1} implies that states, $\vec{s}$, from ${\bf{\ell_p}}$
have a minimum lower bound on
PAR$_l$ (upper bound on LE) when the associated $[n,k,d]$
code, ${\bf{C}}$, satisfies
$k = \lfloor \frac{n}{2} \rfloor$, with PAR$_l \ge 2^{\lceil \frac{n}{2} \rceil}$.
In this case LE coincides with the \index{Schmidt Measure} 'Schmidt Measure', ${\bf{P}}$
\cite{Eis:Ent,Brie:Ent}, defined as $\log_2$ of the minimal number of non-zero entries in any
state, $\vec{s'}$, where $\vec{s'} \equiv_{\begin{tiny} \mbox{LU} \end{tiny}} \vec{s}$.
\begin{thm}
If $\vec{s} \in {\bf{\ell_p}}$ then $\vec{s}$
is LU equivalent to the indicator for an $[n,k,d]$ code, and the Schmidt
Measure, $P$, satisfies,
$ P(\vec{s}) \le \mbox{min}(k,n-k) $.
\label{thmA}
\end{thm}

Here is a stronger result.
\begin{thm}
Let $s({\bf{x}}) =
(-1)^{x_{\pi(0)}x_{\pi(1)} + x_{\pi(1)}x_{\pi(2)} + \ldots + x_{\pi(n-2)}x_{\pi(n-1)}}$,
where $\pi$ is any permutation of the indeces $\{0,1,\ldots,n-1\}$.
Then PAR$_l = 2^{\lceil \frac{n}{2} \rceil}$.
\label{thmx2}
\end{thm}
\begin{proof}
Corollary 6 of Section 7 of \cite{Par:Gol}.
 \hspace{5mm} \mbox{ }\rule{2mm}{3mm} \end{proof}
\cite{Brie:Ent} also considers the "line graph" states of Theorem \ref{thmx2}
(e.g. $|\Phi_4>$ of \cite{Brie:Ent} is
LU equivalent to $(-1)^{x_0x_1 + x_1x_2 + x_2x_3}$). In \cite{Brie:Ent} Persistency of
Entanglement (Section \ref{sec8}) of such states is proven to be $\lfloor \frac{n}{2}
\rfloor$. This is the same as proof of PAR$_l$, because Persistency, LE,
and Schmidt Measure all coincide for these states.
These states also have PAR $\le 2.0$ under the one-dimension complex
Discrete Fourier Transform \cite{Dav:PF,Pat:PF,Par:Gol}

We now show how to compute the PAR of any $HI$ transform of a member of
${\bf{\ell_p}}$.
Let ${\bf{s}} \in {\bf{\ell_p}}$. Recalling Definition \ref{dfa6}, let $k = |{\bf{T_{C^{\perp}}}}|$,
$k^{\perp} = |{\bf{T_C|}}$, and $k + k^{\perp} = n$. Without loss of
generality we renumber integer sets ${\bf{T_{C^{\perp}}}}$ and
${\bf{T_C}}$ so that
${\bf{T_{C^{\perp}}}} = \{0,1,\ldots,k-1\}$ and ${\bf{T_C}} = \{k,k+1,\ldots,n-1\}$.
Let
${\bf{t_{C^{\perp}} \subset T_{C^{\perp}}}}$ and ${\bf{t_C \subset T_C}}$, where
$h = |{\bf{t_{C^{\perp}}}}|$ and $h^{\perp} = |{\bf{t_C|}}$. Let
${\bf{x_{t^{\perp}}}} = \{x_i | i \in {\bf{t_{C^{\perp}}}}\}$,
${\bf{x_t}} = \{x_i | i \in {\bf{t_C}}\}$, and ${\bf{x_*}} = {\bf{x_{t^{\perp}} \cup x_t}}$.
We define ${\bf{M}}$ to be a $k \times k^{\perp}$
binary matrix where $M_{i,j-k} = 1$ iff $x_ix_{j} \in p({\bf{x}})$, and $M_{i,j-k} = 0$
otherwise. Thus $p({\bf{x}}) =
\sum_{i \in {\bf{T_{C^{\perp}}}}} x_i(\sum_{j \in {\bf{T_C}}} M_{i,j-k}x_{j})$.
Then we
define a submatrix, ${\bf{M_t}}$, of ${\bf{M}}$, which comprises only the rows
and columns of ${\bf{M}}$ specified by ${\bf{t_{C^{\perp}}}}$ and
${\bf{t_C}}$. Let $\chi_t$ be the rank of ${\bf{M_t}}$.
\begin{thm}
Let $\vec{s'}$ be the result of the  action of $\prod_{i \in
{\bf{t_{C^{\perp}} \cup t_C}}} H(i)$ on $\vec{s} \in {\bf{\ell_p}}$. Then,
$$ \mbox{PAR}(\vec{s'}) = 2^{h + h^{\perp} - 2\chi_t} $$
\label{thmX1}
\end{thm}
\begin{proof} We observe that PAR$(\vec{s'}) =
\mbox{PAR}(\mbox{WHT}((-1)^{p_t({\bf{x_*}})}))$, where \newline
$p_t({\bf{x_*}}) =
\sum_{i \in {\bf{t_{C^{\perp}}}}} x_i(\sum_{j \in {\bf{t_C}}} M_{i,j-k}x_{j})$.
By affine transformation of the variables in ${\bf{x_*}}$
we can rewrite $p_t({\bf{x_*}})$ as,
$$ p_t({\bf{x_*}}) =
\sum_{i=0}^{\chi_t - 1} f_i({\bf{x_{t^{\perp}}}})g_i({\bf{x_t}}) $$
where $f_i$ and $g_i$ are linearly independent linear combinations of the
variables in ${\bf{x_{t^{\perp}}}}$ and ${\bf{x_t}}$, respectively, $\forall i$. For clarity we further
rewrite $p_t$ as,
$$ p_t({\bf{x_*}}) = p_t({\bf{y}},{\bf{z}}) = \sum_{i=0}^{\chi_t - 1} y_iz_i $$
where the $y_i$ and $z_i$ are linearly independent binary variables over the spaces
${\bf{x_{t^{\perp}}}}$ and ${\bf{x_t}}$, respectively. It is known that the
WHT over the $2\chi_t$ binary variables ${\bf{y \cup z}}$ of $(-1)^{p_t({\bf{y}},{\bf{z}})}$
has PAR $ = 1$, (i.e. $p_t$ is bent, \cite{MacW:Cod}). Therefore the
WHT of $(-1)^{p_t({\bf{x_*}})}$
over the $h + h^{\perp}$ binary variables in ${\bf{x_*}}$
has PAR $ = 2^{h + h^{\perp} - 2\chi_t}$.  \hspace{5mm} \mbox{ }\rule{2mm}{3mm} \end{proof}
\begin{cor}
As $0 \le \chi_t \le \mbox{min}(h,h^{\perp})$, it follows that
$\mbox{PAR}(\vec{s'}) \ge 2^{|h - h^{\perp}|}$
\end{cor}
Theorem \ref{thmx1} also follows trivially from Theorem \ref{thmX1}.

Each ${\bf{M_t}}$ is associated with a different code, $C_{M_t}$, with
dimension $n - \log_2(\mbox{PAR}(\vec{s'}))$. In particular, when
${\bf{t_{C^{\perp}}}} = {\bf{T_{C^{\perp}}}}$ and ${\bf{t_C}} = \emptyset$
then $C_{M_t} = C$ with dimension $k$. Similarly, when
${\bf{t_{C^{\perp}}}} = \emptyset$ and ${\bf{t_C}} = {\bf{T_C}}$
then $C_{M_t} = C^{\perp}$ with dimension $k^{\perp} = n-k$. When
${\bf{t_{C^{\perp}}}} = {\bf{t_C}} = \emptyset$ then $C_{M_t}$ has
dimension $n$.
Each code, $C_{M_t}$, is specified by the bit positions which are acted on
by $H(i)$, i.e. the integers in ${\bf{t_{C^{\perp}}}} \cup {\bf{t_C}}$.
\begin{lem}
Let $\vec{s} \in {\bf{\ell_p}}$, and let $\vec{s'}$ be a $HI$ transform of
$\vec{s}$. Let $\vec{s''} = H(i)[\vec{s'}])$. Then,
$$ \frac{\mbox{PAR}(\vec{s''})}{\mbox{PAR}(\vec{s'})} \in \{\frac{1}{2},2\} $$
\label{lemZ5}
\end{lem}
\begin{proof}
If an integer, $e$, is added or removed to or from either ${\bf{t_{C^{\perp}}}}$ or ${\bf{t_C}}$
then the associated qubit, $x_e$, is
acted on by $H(e)$, and the associated row or column is added or removed
to give ${\bf{M_e}}$, with associated rank, $\chi_e$.
From Theorem \ref{thmX1} we see that if $\chi_e = \chi_t$ or $\chi_t - 1$, respectively,
then PAR doubles whereas, if $\chi_e = \chi_t$ or $\chi_t + 1$, respectively,
then PAR halves.
 \hspace{5mm} \mbox{ }\rule{2mm}{3mm} \end{proof}
\begin{thm}
PAR$_l$ of $\vec{s} \in {\bf{\ell_p}}$ is found in the
$HI$ multispectra of $\vec{s}$.
\label{thmX2}
\end{thm}
\begin{proof} Section \ref{App1}.
 \hspace{5mm} \mbox{ }\rule{2mm}{3mm} \end{proof}
\section{Weight Hierarchy and Stubborness of Entanglement (SE)}
\label{sec8}
\subsection{Weight Hierarchy}
We prove that weight hierarchy of an $[n,k,d]$ linear code, ${\bf{C}}$,
can be obtained from the $HI$ multispectra.
\begin{df}
The Weight Hierarchy of ${\bf{C}}$, is
a series of parameters, $d_j$,
$0 \le j \le k$, representing the smallest blocklength of a linear sub-code of
${\bf{C}}$ of dimension $j$, where $d_k = n$, $d_1 = d$, and $d_0 = 0$.
\label{dfZ9}
\end{df}
\begin{thm}
Let $\vec{s_c}$ be the indicator of an $[n,k,d]$ binary linear error-correcting
code, ${\bf{C}}$.
Let ${\bf{Q}} \subset \{0,1,\ldots,n-1\}$. Let,
\begin{equation}
m_{\bf{Q}} = \frac{|{\bf{Q}}| + \log_2(\mu) - n  + k}{2}, \hspace{3mm}
\mbox{ where } \mu = \mbox{PAR}(\vec{s_c'})
\label{eqW1}
\end{equation}
and $\vec{s_c'} = \prod_{t \in {\bf{Q}}} H(t)[\vec{s_c}]$.
Then the Weight Hierarchy of ${\bf{C}}$ is found from the $HI$
multispectra of $\vec{s_c}$, where,
$$ d_j = \mbox{min}_{{\bf{Q}} | m_Q = j}(|{\bf{Q}}|) $$
\label{thmQ}
\end{thm}
\begin{proof} Section \ref{App2} \hspace{5mm} \mbox{ }\rule{2mm}{3mm} \end{proof}
\subsection{Stubborness of Entanglement}
\begin{thm}
Let $\vec{s} \in {\bf{\ell_p}}$. A single qubit measurement on $\vec{s}$
gives $\vec{s'}$, and has one of two results:
\begin{itemize}
	\item Destructive Measurement, where
$|\{s'_i | s'_i \ne 0\}| = \frac{1}{2}|\{s_i | s_i \ne 0\}|$, and
PAR$(\vec{s'}) = 2\mbox{PAR}(\vec{s})$.
	\item Redundant Measurement, where
$|\{s'_i | s'_i \ne 0\}| = |\{s_i | s_i \ne 0\}|$, and
PAR$(\vec{s'}) = \mbox{PAR}(\vec{s})$.
\end{itemize}
where $|s_i|,|s_i'| \in \{0,1\}$.
\label{thmW}
\end{thm}
\begin{proof} Section \ref{App2}.
\hspace{5mm} \mbox{ }\rule{2mm}{3mm} \end{proof}
Let the entanglement order of a system be the size (in qubits) of the largest entangled
subsystem of the system.
\begin{df}
A most-destructive series of $j$ single-qubit measurement over some set of possible
measurements on $\vec{s}$ produces a final state $\vec{s'}$ such that
$\mbox{entanglement order}(\vec{s}) - \mbox{entanglement order}(\vec{s'})$ is maximised.
\label{dfZ4}
\end{df}
\begin{df}
The Stubborness of Entanglement (SE) is a series of parameters, $\beta_j$,
$0 \le j \le k'$, representing the smallest possible entanglement order, $\beta_j$,
after $k'-j$ most-destructive measurements of an $n$-qubit system, where
$\beta_{k'} = n$, $\beta_0 = 0$.
\label{dfXM3}
\end{df}
Note that 'Persistency of Entanglement', as defined in \cite{Brie:Ent}, is $k'$, as it
is the minimum number of measurements necessary
to reduce the state to entanglement order 0. This section shows that the similarity
between Definitions \ref{dfZ9} and \ref{dfXM3} implies that the Weight Hierarchy
upper bounds LE (Corollary \ref{dfXM1}) and becomes {\underline{equivalence}}
for optimal or near-optimal binary linear ECCs (Theorem \ref{thmX4}).
For a state from ${\bf{\ell_p}}$ which is LU-equivalent to an $[n,k,d]$ binary
linear ECC, we show that the minimum number of
single qubit measurements, {\bf{in any basis}}, required to completely disentangle such
a state is $\le k$ (Theorem \ref{thma7}).
We also show that, for such states, the series of 'most-destructive'
measurements occurs in the $HI$ multispectra (Theorem \ref{thmX3}).
\begin{df}
The ${\bf{C}}$-basis for $\vec{s}$ is the basis in which the vector
becomes an indicator for the code ${\bf{C}}$.
This indicator has
non-negative, multi-valued output and, for $\vec{s} \in {\bf{\ell_p}}$,
the indicator has binary values.
Without loss of generality, we always assume
$|{\bf{C}}| \le |{\bf{C^{\perp}}}|$. There is not a ${\bf{C}}$-basis for every
$\vec{s}$ but, for $\vec{s} \in {\bf{\ell_p}}$, there exists a
${\bf{C}}$-basis and ${\bf{C^{\perp}}}$-basis.
\label{dfa9}
\end{df}
\begin{small}
{\bf{Example: }} By Theorem \ref{thma2},
the bipartite bipolar quadratic state \newline
$s({\bf{x}}) = (-1)^{x_0x_1 + x_1x_2 + x_2x_3 + x_3x_0}$
is LU equivalent to $s_c({\bf{x}}) = (x_0 + x_1 + x_3 + 1)(x_1 + x_2 + x_3 + 1)$ which is the indicator
for ${\bf{C}} = \{0000,0111,1010,1101\}$. Measuring qubit 0 of $s({\bf{x}})$ in the ${\bf{C}}$-basis
gives the indicator for the subcode $\{000,111\}$ if $x_0 = 0$, and the subcode $\{010,101\}$ if
$x_0 = 1$.
\end{small}
\begin{lem}
Let $\vec{s} \in {\bf{\ell_p}}$. Let $\vec{s'}$ be a $HI$ transform of $\vec{s}$.
Then $n - \log_2(\mbox{PAR}(\vec{s'}))$ destructive measurements
in the $\vec{s'}$-basis are sufficient to completely destroy entanglement in $\vec{s}$.
\label{lemZ9}
\end{lem}
\begin{proof}
Follows from Theorem \ref{thmW}.
 \mbox{ }\rule{2mm}{3mm} \end{proof}
\begin{thm}
Let $\vec{s} \in {\bf{\ell_p}}$. Then $\mbox{LE}(\vec{s})$ measurements is the
minimum number of measurements required to completely destroy the entanglement
of $\vec{s}$.
\label{thmXM1}
\end{thm}
\begin{proof}
From Definition \ref{dfXM2}, Theorem \ref{thmX2}, and Lemma \ref{lemZ9}.
 \hspace{5mm} \mbox{ }\rule{2mm}{3mm} \end{proof}
\begin{thm}
Let $k$ and $k'$ be as defined in Definitions \ref{dfZ9} and \ref{dfXM3}.
Then $k' \le k$. In words,
$k$ destructive measurements in the ${\bf{C}}$-basis suffice to destroy all
entanglement in $\vec{s}$, where
$\vec{s} \in {\bf{\ell_p}}$ is equivalent to the binary linear $[n,k,d]$
error-correcting code, ${\bf{C}}$.
\label{thma7}
\end{thm}
\begin{proof}
From Theorem \ref{thmx1} and Lemma \ref{lemZ9}.
 \hspace{5mm} \mbox{ }\rule{2mm}{3mm} \end{proof}
\begin{thm}
$k-j$ destructive measurements in the ${\bf{C}}$-basis suffice to reduce
the entanglement order of $\vec{s}$ to $d_j$ qubits, where
$\vec{s} \in {\bf{\ell_p}}$ is equivalent to the binary linear $[n,k,d]$
error-correcting code, ${\bf{C}}$.
\label{thma9}
\end{thm}
\begin{proof}
By recursive application of the proof for Theorem \ref{thma7}.
 \hspace{5mm} \mbox{ }\rule{2mm}{3mm} \end{proof}
\begin{cor}
For ${\bf{s}} \in {\bf{\ell_p}}$,
$$ k' < k \hspace{5mm} \mbox{ or } \hspace{5mm} \beta_j \le d_j \hspace{2mm} \mbox{ if }
k' = k $$
In words, for states from ${\bf{\ell_p}}$, the Weight Hierarchy is an upper bound on
Stubborness of Entanglement.
\label{corXM1}
\end{cor}
\begin{thm}
The series of residual entanglement orders, $\beta_j'$, resulting
from $j$ most-destructive measurements of $\vec{s_c}$ in the $\vec{s_c}$-basis,
satisfies $\beta_j' = d_j$. In words, the Stubborness of Entanglement
confined to the $\vec{s_c}$-basis is equivalent to the weight hierarchy of ${\bf{C}}$.
\label{thmQ1}
\end{thm}
\begin{proof}
By recursive application of Theorems \ref{thmW}, \ref{thmQ}, Lemma \ref{lemZ9},
and Theorem \ref{thmXM1}.
 \mbox{ }\rule{2mm}{3mm} \end{proof}

{\bf{Warning: }} Let ${\bf{S^c}}_j$ be a size $d_j$ subset of $\{0,1,\ldots,n-1\}$,
such that the linear subcode of ${\bf{C}}$ defined by ${\bf{S^c}}_j$ has dimension
$j$. Then it is {\underline{not}} necessarily the case that
${\bf{S^c}}_j \subset {\bf{S^c}}_{j+1}$ for any $j$.
\begin{df}
If $\vec{s_c}$ is a binary indicator for ${\bf{C}}$
such that ${\bf{S^c_{j} \subset S^c_{j+1}}}$
$\forall j$, then $\vec{s}$ satisfies the 'chain condition'. The weaker definition
of 'greedy weights' identifies only weights $d_j$ for which
${\bf{S^c_{j} \subset S^c_{j+1}}}$.
\label{dfa10}
\end{df}
Similarly, let ${\bf{S^q}}_j$ be a size $\beta_j$ subset of $\{x_0,x_1,\ldots,x_{n-1}\}$,
comprising the subset of qubits with maximum entanglement order, $\beta_j$,
after $k'-j$ most-destructive measurements.
Then it is {\underline{not}} necessarily the case that
${\bf{S^q}}_j \subset {\bf{S^q}}_{j+1}$ for any $j$.
\begin{df}
If $\vec{s}$ is such that ${\bf{S^q_{j} \subset S^q_{j+1}}}$
$\forall j$, then $\vec{s}$ satisfies the 'quantum chain condition'.
The weaker definition
of 'quantum greedy weights' identifies only weights $\beta_j$ for which
${\bf{S^q_{j} \subset S^q_{j+1}}}$.
\label{dfb10}
\end{df}
We leave investigation of the quantum chain condition to future work.
\begin{thm}
Let $\vec{s} \in {\bf{\ell_p}}$. Then the Stubborness of Entanglement
of $\vec{s}$ can be computed from a
set of most-destructive measurements in the $HI$-basis of $\vec{s}$.
\label{thmX3}
\end{thm}
\begin{proof} A single-qubit measurement reduces (classical) entropy of
the system by a maximum of 1 bit per measurement. This corresponds to measuring the
qubit value as $0$ or $1$ with equal likelihood ($-(\frac{1}{2}\log_2(\frac{1}{2}) +
\frac{1}{2}\log_2(\frac{1}{2})) = 1$). Destructive measurements in the $HI$-basis of a
state $\vec{s}$ from ${\bf{\ell_p}}$ are always of this form (Theorem \ref{thmW}).
For states from ${\bf{\ell_p}}$, entropy is LE. The theorem follows from
Theorem \ref{thmXM1}.
  \hspace{5mm} \mbox{ }\rule{2mm}{3mm} \end{proof}
\begin{df}
A binary linear $[n,k,d]$ code is here considered optimal for a given $n$ if
$d_1,d_2,\ldots,d_{k-1}$ are maximised for each $k$, $k \le \lfloor \frac{n}{2} \rfloor$,
with priority to the weights, $d_i$, with lowest $i$.
\label{dfZ11}
\end{df}
\begin{thm}
Let $\vec{s} \in {\bf{\ell_p}}$ where $\vec{s}$ is
LU equivalent to an optimal or near-optimal binary linear code of dimension $\le \frac{n}{2}$.
Then Stubborness of Entanglement is equal to the Weight Hierarchy of the code.
\label{thmX4}
\end{thm}
\begin{proof} We require the following Lemma.
\begin{lem}
Let $\vec{s}$ be LU equivalent to an optimal or near-optimal binary linear code, $C$.
If $C$ has dimension $k = \frac{n}{2}$, the $HI$ multispectra of
$\vec{s}$ only has PAR as high as $2^{n-k}$ for two $HI$ transforms, one the WHT
of the other. These two transform spectra, $\vec{s_c}$ and
$\vec{s_{c^{\perp}}}$ are binary indicators for $C$ and $C^{\perp}$,
respectively.
Similarly, when $C$ has dimension $k < \frac{n}{2}$, the $HI$ multispectra of
$\vec{s}$ only has PAR as high as $2^{n-k}$ for one $HI$ transform. This
transform spectra, $\vec{s_c}$, is the binary indicator for $C$.
\label{lemX1}
\end{lem}
\begin{proof} (Lemma \ref{lemX1}).
By Theorem \ref{thmQ}, an optimal or near-optimal
binary linear code should have PAR as low as possible in the $HI$ multispectra.
It is easy to validate Lemma \ref{lemX1} for
$1 \le n \le 4$ by exhaustive computation. Now consider a state, $\vec{s}$,
that satisfies Lemma \ref{lemX1} for $n$ even. Then we can identify (at least)
one $HI$ transform of $\vec{s}$ which gives a PAR of $2^{n-k}$. Let
us call this transform $R$. Let ${\bf{s}} = (-1)^{{\bf{p}}}$.
Let $x_ix_j$ be an arbitrary
term in $\vec{p}$ such that $R$ contains $H(i)$.
Let us increase $n$ by one to $n' = n + 1$ by appending
$x_lx_i$, to create $p^+({\bf{x}}) = p({\bf{x}}) + x_lx_i$,
where $x_l$ is a new qubit added to the system. Then
$R \otimes I_l$ is the only $HI$ transform of
$\vec{s^+} = (-1)^{\vec{p^+}}$ which achieves a PAR as high as $2^{n-k+1}$ so Lemma \ref{lemX1}
is satisfied for $n$ odd, as long as it is satisfied for $n$ even.
Let us now increase $n$ to $n'' = n + 2$ by appending
$x_lx_i + x_mx_j + x_lx_m$, to create $p^{++}({\bf{x}}) =
p({\bf{x}}) + x_lx_i + x_mx_j + x_lx_m$,
where $x_l$ and $x_m$ are new qubits added to the system. Then
$R \otimes I_l \otimes H_m$, $R \otimes H_l \otimes H_m$, and
$R \otimes H_l \otimes I_m$ are $HI$ transforms of
$\vec{s^{++}} = (-1)^{\vec{p^{++}}}$ which achieve a PAR of $2^{n-k+1}$, $2^{n-k}$,
and $2^{n-k-1}$, respectively.
Therefore Lemma \ref{lemX1}
is satisfied for $n$ even as no other $HI$ transform can reach a PAR of $2^{n-k+1}$.
By induction on $n$ it follows that we can
construct states $\vec{s}$ which satisfy the conditions of Lemma
\ref{lemX1} for all $n$ and $k \le \frac{n}{2}$.
 \hspace{5mm} \mbox{ }\rule{2mm}{3mm} \end{proof}
The most-destructive series of measurements on $\vec{s}$ must be performed in the
basis which is the $HI$ transform, $\vec{s'}$, of $\vec{s}$ that achieves highest PAR
(Lemma \ref{lemZ9}, Theorem \ref{thmXM1}).
But, from Lemma \ref{lemX1}, $\vec{s'} = \vec{s_c}$ and,
for $k = \frac{n}{2}$, $\vec{s'} = \vec{s_{c^{\perp}}}$ also. The Theorem
follows from Theorem \ref{thmQ1}.
 \hspace{5mm} \mbox{ }\rule{2mm}{3mm} \end{proof}
\begin{cor}
Quantum states from ${\bf{\ell_p}}$ which have optimum Linear Entanglement and
optimum Stubborness of Entanglement are LU equivalent to binary linear codes
with optimum Weight Hierarchy.
\label{corX2}
\end{cor}
\subsection{Examples}
We abbreviate tensor product expressions such as
${\bf{H}} \otimes {\bf{I}} \otimes {\bf{H}} \otimes {\bf{I}}$ to $HIHI$.
Let $\vec{s_c}$ be the indicator for the code, ${\bf{C}}$, and
LU-equivalent to $\vec{s}$.
Let $\vec{s'_c}$ be the spectrum of $\vec{s_c}$ after application of
$\prod_{i \in {\bf{Q}}} H(i)$, where ${\bf{Q}} \subset \{0,1,\ldots,n-1\}$.
Let $\vec{s} \in {\bf{\ell_p}}$.
Let $s({\bf{x}}) = (-1)^{x_3x_0 + x_0x_2 + x_2x_1 + x_1x_4 + x_4x_0}$, i.e.
$\vec{s} \in {\bf{\ell_p}}$. Then under
$IIHHH$ we get $s_c({\bf{x}}) = (x_0 + x_3 + 1)(x_0 + x_1 + x_2 + 1)(x_0 + x_1 + x_4 + 1)$
(Theorem \ref{thma2}), which is the binary
indicator for the $[5,2,3]$ code, ${\bf{C}} = $ \begin{small}
$\{00000,11010,01101,10111\}$ \end{small}. The action of the $HI$ transform
on $\vec{s_c}$ can be computed from Theorem \ref{thmX1} and
gives the following PARs for the spectrum, $\vec{s'_c}$:
\begin{small} $$ \begin{array}{c|cccccccc}
x_0x_1 : x_2x_3x_4 & III & HII & IHI & HHI & IIH & HIH & IHH & HHH \\ \hline
II & 8 & 4 & 4 & 2 & 4 & 2 & 2 & 1 \\
HI & 4 & 2 & 2 & 1 & 2 & 1 & 1 & 2 \\
IH & 4 & 2 & 2 & 1 & 2 & 4 & 1 & 2 \\
HH & 2 & 1 & 4 & 2 & 1 & 2 & 2 & 4
\end{array} $$ \end{small}
For instance, the 3rd row, second column shows that the
spectrum after application of $IHHII$ on $\vec{s_c}$ has a PAR of $2$, where ${\bf{Q}} = \{1,2\}$.
Similarly, the 2nd row and 6th column shows that the
spectrum after application of $HIHIH$ on $\vec{s_c}$ has a PAR of $1$, where ${\bf{Q}} = \{0,2,4\}$. The
entry for ${\bf{Q}} = \{-\}$ is $2^{n-k} = 8$, the entry for
${\bf{Q}} = \{0,1,2,3,4\}$ is $2^k = 4$, and the entry for ${\bf{Q}} = \{2,3,4\}$ is $1$,
as this is the bipolar form, $\vec{s}$. From the table and by
Theorem \ref{thmX2} we see that PAR$_l = 8$, where the maximum PAR occurs
at $\vec{s_c}$. Therefore, by Theorem \ref{thmX4}, we can equate SE with
Weight Hierarchy.
We now determine the weight hierarchy for this
$[5,2,3]$ code from a most-destructive PAR trajectory in the $HI$ multispectra,
using Theorem \ref{thmQ}. Before
measurement of $\vec{s_c}$ we have ${\bf{Q}} = \{0,1,2,3,4\}$ and, after $k = 2$
destructive measurements in the ${\bf{C}}$-basis we have ${\bf{Q}} = \{-\}$.
A series of most-destructive series of measurements achieves a system
with smallest entanglement order after the smallest number of measurements.
Here is a
PAR trajectory corresponding to a most-destructive series of measurements in the ${\bf{C}}$-basis
where, without
loss of generality, the measurement result is always 0. A 'freed'
qubit is the result of redundant measurement.
(For states from ${\bf{\ell_p}}$
the optimal measurement strategy is independent of measurement outcome. This is not generally true for other
quantum states):
\begin{small}
$$ \begin{array}{l|ccccccccccc}
\mbox{action}    & \mbox{WHT$(s_c({\bf{x}}))$} & & \mbox{measure} & & \mbox{free} & & \mbox{measure} & & \mbox{free} & & \mbox{free} \\
         & & & \mbox{qubit } 2 & & \mbox{qubit } 4 & & \mbox{qubit } 3 & &  \mbox{qubit } 1
         & & \mbox{qubit } 0 \\
{\bf{Q}} & \{0,1,2,3,4\} & & \{0,1,3,4\} & & \{0,1,3\} & & \{0,1\} & & \{0\} & & \{-\} \\
$HI$     & HHHHH  & \rightarrow   & HHIHH   & \rightarrow   & HHIHI   & \rightarrow   &
     HHIII   & \rightarrow   & HIIII   & \rightarrow & IIIII \\
\mbox{PAR} & 4 & & 2 & & 4 & & 2 & & 4 & & 8 \\
{\bf{C}} &
\begin{array}{c}
00000 \\ 11010 \\ 01101 \\ 10111
\end{array} & &
\begin{array}{c}
00\_00 \\ 11\_10
\end{array} & &
\begin{array}{c}
00\_0\_ \\ 11\_1\_
\end{array} & &
\begin{array}{c}
00\_\_\_
\end{array} & &
\begin{array}{c}
0\_\_\_\_
\end{array} & &
\begin{array}{c}
\_\_\_\_\_
\end{array} \\
m_{\bf{Q}} & 2 & & 1 & & 1 & & 0 & & 0 & & 0 \\
d_j & d_2 = 5 & &   & & d_1 = 3 & & & & & & d_0 = 0
\end{array} $$
\end{small}
The PAR trajectory shown above gives the weight hierarchy,
$d_2 = 5$, $d_1 = 3$, $d_0 = 0$, where $d_1$ is the Hamming Distance of ${\bf{C}}$,
which, in this case, is also SE, where $\beta_i = d_i$.

\begin{figure}
\begin{center}
\includegraphics[width=.4\textwidth]{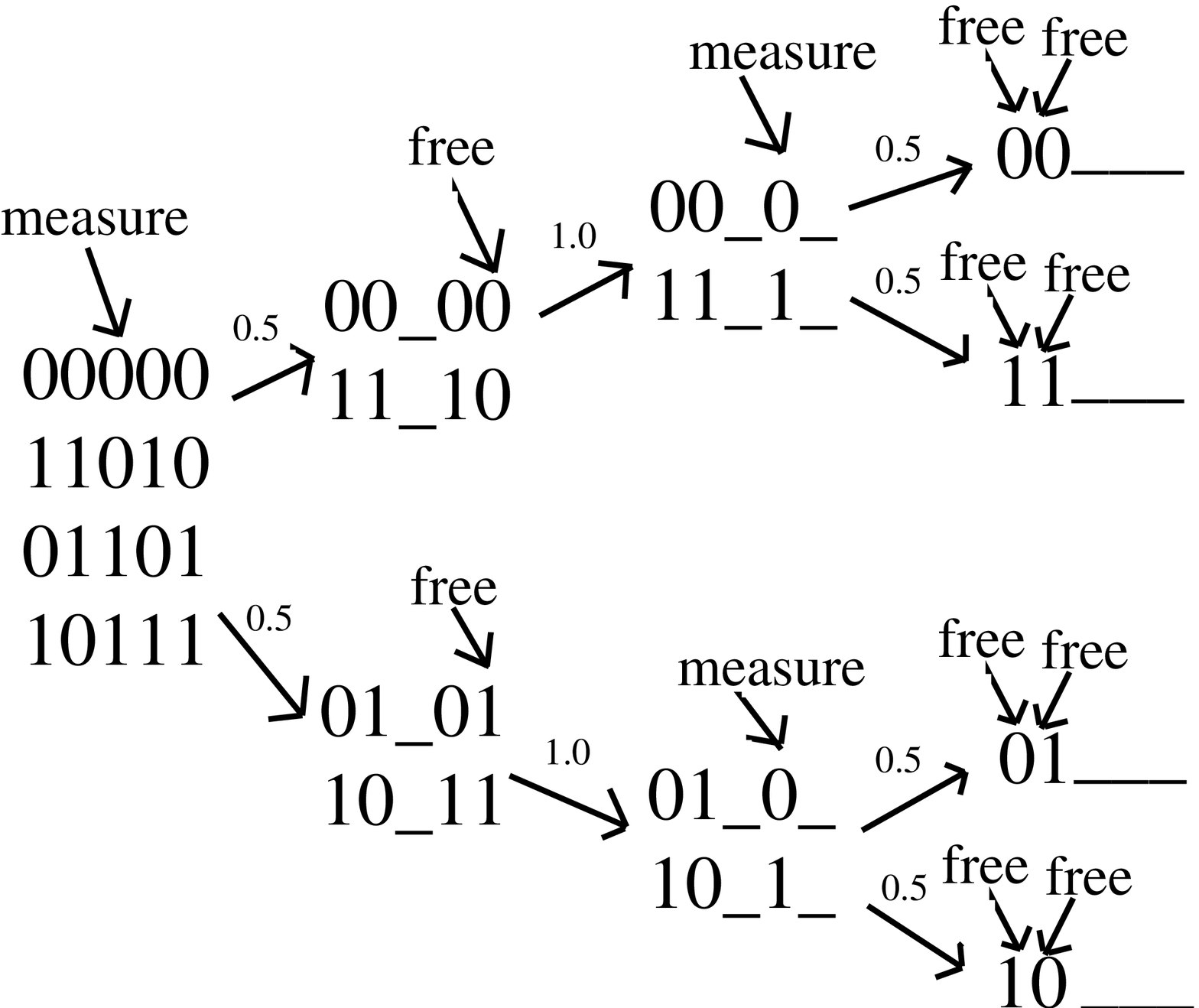} \hspace{5mm} \mbox{ } \hspace{5mm} \mbox{ }
\includegraphics[width=.4\textwidth]{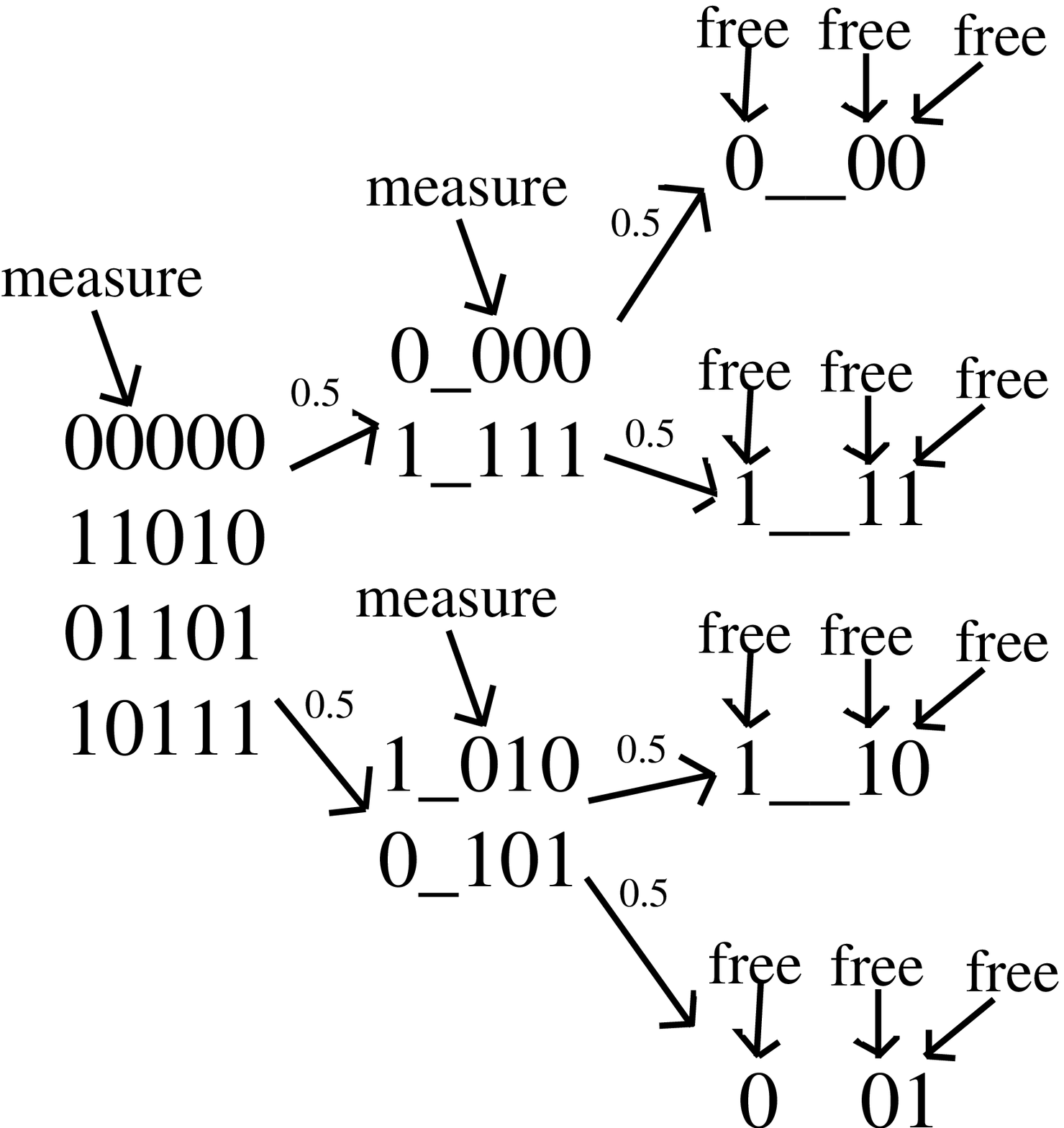}
\caption[]{Most-Destructive (lh) and Least-Destructive (rh) Measurement in the ${\bf{C}}$-Basis \label{eps8}}
\end{center}
\end{figure}

\begin{figure}
\begin{center}
\includegraphics[width=1.1\textwidth]{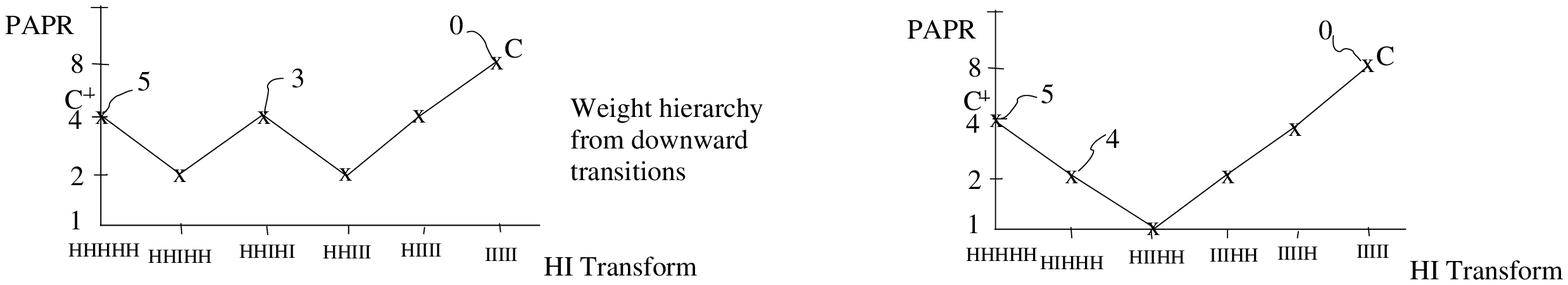}
\caption[]{PAR Trajectories for Most-Destructive (lh) and
Least-Destructive (rh) ${\bf{C}}$-Basis Measurements \label{eps9}}
\end{center}
\end{figure}

In contrast, here is a least destructive set of measurements in the ${\bf{C}}$-basis for
the same example state, $s({\bf{x}}) = (-1)^{x_3x_0 + x_0x_2 + x_2x_1 + x_1x_4 +
x_4x_0}$.
\begin{small}
$$ \begin{array}{l|ccccccccccc}
\mbox{action}    & \mbox{WHT$(s_c({\bf{x}}))$} & & \mbox{measure} & & \mbox{measure} & & \mbox{free} & & \mbox{free} & & \mbox{free} \\
         & & & \mbox{qubit } 2 & & \mbox{qubit } 4 & & \mbox{qubit } 3 & &  \mbox{qubit } 1
         & & \mbox{qubit } 0 \\
{\bf{Q}} & \{0,1,2,3,4\} & & \{0,2,3,4\} & & \{0,3,4\} & & \{3,4\} & & \{4\} & & \{-\} \\
$HI$     & HHHHH  & \rightarrow   & HIHHH   & \rightarrow   & HIIHH   & \rightarrow   &
     IIIHH   & \rightarrow   & IIIIH   & \rightarrow & IIIII \\
\mbox{PAR} & 4 & & 2 & & 1 & & 2 & & 4 & & 8 \\
{\bf{C}} &
\begin{array}{c}
00000 \\ 11010 \\ 01101 \\ 10111
\end{array} & &
\begin{array}{c}
0\_000 \\ 1\_111
\end{array} & &
\begin{array}{c}
00\_\_00
\end{array} & &
\begin{array}{c}
\_\_\_00
\end{array} & &
\begin{array}{c}
\_\_\_\_0
\end{array} & &
\begin{array}{c}
\_\_\_\_\_
\end{array} \\
m_{\bf{Q}} & 2 & & 1 & & 0 & & 0 & & 0 & & 0
\end{array} $$
\end{small}
Fig \ref{eps8} shows the two measurement scenarios described above, and Fig
\ref{eps9} shows their corresponding PAR trajectories.
The weight hierarchy corresponds to the highest possible
PAR trajectory.
For more general quantum states,
states requiring most measurements to destroy entanglement will have flattest
LU multispectra.

\begin{figure}
\begin{center}
\includegraphics[width=.3\textwidth]{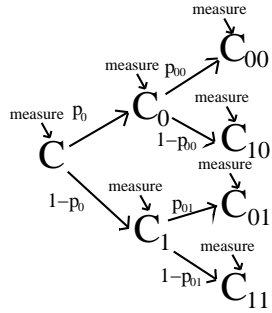}
\caption[]{A General Measurement Trajectory \label{eps10}}
\end{center}
\end{figure}

For quantum states outside ${\bf{\ell_p}}$
the most destructive set of measurements generally occurs outside the $HI$ multispectra,
but it is hoped that further research will lead to an identification of the
spectral location of this most-destructive set of measurements.
For more general states, the strategy
becomes dependent on measurement outcomes. The entropy measure, $-p_0\log_2(p_0) - p_1\log_2(p_1)$, where
$p_1 = (1 - p_0)$, is then
the natural generalisation of $k$ measurements, as shown in Fig \ref{eps10}, giving rise
to 'Entropic Weight Hierarchy'. (In this context, 'freed' qubits contribute zero to the
entropy sum as $-p_0\log_2(p_0) - p_1\log_2(p_1)$ is zero if $p_0$ or $p_1 = 0$).
Entropic Weight Hierarchies will be pursued in future papers.
\section{Cryptographic Measures of Entanglement}
\label{sec8a}
\subsection{An Upper Bound on PAR$_l$}
We now consider properties of a bipolar sequence, $\vec{s}$, that are familiar to a
cryptanalyst. Let $\vec{s_H}$ be the WHT of $\vec{s}$.
\begin{df}
The Nonlinear Order ($N$) of $\vec{s}$ is given by,
$$ N(\vec{s}) = n - \log_2(\mbox{PAR}(\vec{s_H})) $$
\label{dfx1}
\end{df}
which, by Definition \ref{dfXM2}, is immediately a trivial upper bound on LE.
For $\vec{s}$ linear, PAR $ = 2^n$ and $N = 0$, and for $\vec{s}$ as nonlinear as possible,
PAR $ = 1$, and
$N = n$. Nonlinear order is therefore a PAR measure in the WHT
domain. A second measure commonly used is 'Correlation-Immunity Order' (CI$-t$)
\cite{Xia:CI}.
\begin{df}
$\vec{s}$ has Correlation-Immunity Order $t$ (CI-$t$) if
$\vec{s_H}$ is zero at all indeces with binary weight $> 0$ and $\le t$.
\label{dfx2}
\end{df}
Note that, if $\vec{s_H}$ is an indicator for a code, ${\bf{C_H}}$, with
minimum Hamming Distance $d = t + 1$, then ${\bf{s}}$ has CI-$t$.
\footnote{There is also an entropic version of CI-$t$, related to Fig \ref{eps10}, but we
will investigate this more general parameter in future work. Entanglement
measures discussed in this paper, such as spectral peak, spectral weight, and spectral entropy,
can all be described by the 'Renyi Entropy' \cite{Zyc:Ent}}.
The Nonlinear Order and Correlation-Immunity of a state from ${\bf{\ell_p}}$ provide an upper
and lower bound on PAR$_l$.
\begin{thm}
Let $s({\bf{x}}) = (-1)^{p({\bf{x}})} \in {\bf{\ell_p}}$ and
have nonlinear order, $N$. Then,
$$ n-N \le \log_2(\mbox{PAR}_l(\vec{s})) \le n-\frac{N}{2} $$
Furthermore, if the Correlation-Immunity of $s$ is CI-$t$,
$$ n-N \le \log_2(\mbox{PAR}_l(\vec{s})) \le \mbox{max}(n-t-1-\frac{N}{2},n-N) $$
where $0 \le t + 1 \le n - N $.
\label{thmX5}
\end{thm}
\begin{proof}
The lower bound follows from Definitions \ref{dfXM2} and \ref{dfx1}. For the upper bound,
the application of $H(i)$ on a member of ${\bf{\ell_p}}$ either decreases or increases
PAR by a factor of 2 (Lemma \ref{lemZ5}).
Therefore, PAR$(\prod_{i \in {\bf{T}}}H(i)[\vec{s}])$ is
maximised for some subset, ${\bf{T}}$, if $\exists {\bf{T'}}$ such that,
\begin{equation} \mbox{PAR}(\prod_{i \in {\bf{T'}}}H(i)[\vec{s}]) =
2^{|{\bf{T}}| - | \hspace{1mm}  |{\bf{T}}| - |{\bf{T'}}| \hspace{1mm} |} \label{eqXXX3} \end{equation}
for at least one subset, ${\bf{T'}}$, of each size $i$, $0 \le i \le n$.
But, for $|{\bf{T'}}| = n$,
$ \prod_{i \in {\bf{T'}}} H(i)[\vec{s}] = \vec{s_H} $,
where $\vec{s_H} = \mbox{WHT}(\vec{s})$. We have,
\begin{equation} \mbox{PAR}(\vec{s_H}) = 2^{n-N} \label{eqXXX1} \end{equation}
Combining (\ref{eqXXX3}) and (\ref{eqXXX1}) gives,
$$ 2^{2|{\bf{T}}| - n} = 2^{n-N} \Rightarrow \mbox{PAR}_l(\vec{s}) \le 2^{|{\bf{T}}|} = 2^{n - \frac{N}{2}} $$
thereby satisfying the first part of the Theorem.
Now, if $\vec{s}$ is also known to be CI-$t$, then this is equivalent to saying that,
\begin{equation} \mbox{PAR}(\prod_{i \in {\bf{T'}}}H(i)[\vec{s}]) = 2^{n - N - t - 1}
\label{eqXXX2}\end{equation}
$\forall {\bf{T'}}$ where $|{\bf{T'}}| = n - t - 1$.
Combining (\ref{eqXXX3}) and (\ref{eqXXX2}) gives,
$$ \begin{array}{ll}
\mbox{PAR}_l(\vec{s}) \le 2^{|{\bf{T}}|} = 2^{n - t - 1 - \frac{N}{2}}, & \hspace{10mm} 0 \le |{\bf{T}}| \le
n-t-1 \\
\mbox{PAR}_l(\vec{s}) \le 2^{n-N}, & \hspace{10mm} n-t \le |{\bf{T}}| \le n
\end{array} $$
The Theorem follows.
 \hspace{5mm} \mbox{ }\rule{2mm}{3mm} \end{proof}
\subsection{Example}
$s({\bf{x}}) = (-1)^{x_0x_1 + x_0x_3 + x_0x_5 + x_1x_2 + x_1x_4 + x_2x_3 + x_2x_5 + x_3x_4 + x_4x_5}$.
Then $\vec{s} \in {\bf{\ell_p}}$ and is LU equivalent, via $IHIHIH$, to $\vec{s_c}$,
which is the binary
indicator for a $[6,3,2]$ binary linear ECC. As $k = 3$,
no more than $3$ measurements in the ${\bf{C}}$-basis are required to completely destroy the entanglement of
$\vec{s}$ (Theorem \ref{thma7}). Moreover, PAR$_l(\vec{s}) \ge 2^3$ (Theorem \ref{thmx1}).
Alternatively, we could examine $\vec{s}$ from a cryptographic angle. $\vec{s}$ is LU equivalent, via
$HHHHHH$ (WHT) to $\vec{s_H}$, where
\begin{tiny} $\vec{s_H} = (+00000000000000000000+00000000000000000000+00000000000000000000-)$ \end{tiny},
where '$+$' means $1$ and '$-$' means $-1$.
$\vec{s_H}$ is an indicator for the $[6,2,3]$ ECC, ${\bf{C_H}} = $
\begin{small} $\{000000,010101,101010,111111\}$ \end{small}
(ignoring phase). So, in the 'WHT-basis', only
2 measurements are required to completely destroy the entanglement of $\vec{s}$,
PAR$_l(\vec{s}) \ge 2^4$
(nonlinear order is 2), and $\vec{s}$ is CI$-1$, (Hamming Distance is 2).
We further examine PARs of all $HI$-multispectra of $\vec{s}$ to get the following table ({\bf{relative to
$\vec{s_c}$}}):
\begin{small} $$ \begin{array}{c|cccccccc}
x_0x_1x_2 : x_3x_4x_5 & III & HII & IHI & HHI & IIH & HIH & IHH & HHH \\ \hline
III & 8 & 4 & 4 & 2 & 4 & 2 & 2 & 1 \\
HII & 4 & 2 & 8 & 4 & 2 & 1 & 4 & 2 \\
IHI & 4 & 2 & 2 & 1 & 2 & 1 & 1 & 2 \\
HHI & 2 & 1 & 4 & 2 & 1 & 2 & 2 & 4
\\
IIH & 4 & 2 & 8 & 4 & 2 & 1 & 4 & 2 \\
HIH & 8 & 4 & 16 & 8 & 4 & 2 & 8 & 4 \\
IHH & 2 & 1 & 4 & 2 & 1 & 2 & 2 & 4 \\
HHH & 4 & 2 & 8 & 4 & 2 & 4 & 4 & 8
\end{array} $$ \end{small}
There is only one PAR of 16 in the table, occuring at $HIHIHI$ relative to $\vec{s_c}$, which is
$HHHHHH$ relative to $\vec{s}$. So, by Theorem \ref{thmX2},
$\vec{s}$ has PAR$_l = 16$
and, by Theorem \ref{thmX3}, SE (measured in the WHT-basis) of $\beta_2 = 6,\beta_1 = 3,\beta_0 = 0$, as shown below.
\begin{tiny}
$$ \begin{array}{l|ccccccccccccc}
\mbox{action}    & s({\bf{x}}) & & \mbox{measure} & & \mbox{free} & & \mbox{free} & & \mbox{measure} & & \mbox{free} & & \mbox{free} \\
         & & & \mbox{qubit } 0 & & \mbox{qubit } 2 & & \mbox{qubit } 4 & &  \mbox{qubit } 1
         & & \mbox{qubit } 3 & & \mbox{qubit } 5 \\
{\bf{Q}} & \{0,1,2,3,4,5\} & & \{1,2,3,4,5\} & & \{1,3,4,5\} & & \{1,3,5\} & & \{3,5\} & & \{5\} & & \{-\} \\
$HI$     & HHHHHH  & \rightarrow   & IHHHHH   & \rightarrow   & IHIHHH   & \rightarrow   &
     IHIHIH   & \rightarrow   & IIIHIH   & \rightarrow & IIIIIH & \rightarrow & IIIIII \\
\mbox{PAR} & 1 & & 2 & & 4 & & 8 & & 4 & & 8 & & 16 \\
{\bf{C}} &
\begin{array}{c}
000000 \\ 010101 \\ 101010 \\ 111111
\end{array} & &
\begin{array}{c}
\_00000 \\ \_10101
\end{array} & &
\begin{array}{c}
\_0\_000 \\ \_1\_101
\end{array} & &
\begin{array}{c}
\_0\_0\_0 \\ \_1\_1\_1
\end{array} & &
\begin{array}{c}
\_\_\_0\_0
\end{array} & &
\begin{array}{c}
\_\_\_\_\_0
\end{array} & &
\begin{array}{c}
\_\_\_\_\_\_
\end{array} \\
m_{\bf{Q}} & 2 & & 1 & & 1 & & 1 & & 0 & & 0 & & 0 \\
\beta_j & \beta_2 = 6 & &   & &  & & \beta_1 = 3 & & & & & & \beta_0 = 0
\end{array} $$
\end{tiny}
We find that $N = 2$ and $t = \beta_1 = 3$ so, by Theorem \ref{thmX5}, it follows that
$$ 4 \le \mbox{PAR}_l \le \mbox{max}(1,4) $$
which agrees exactly with computation, as PAR$_l = 4$.
\section{Non-Bipartite Quadratic Bipolar States}
\label{sec9}
\begin{figure}
\begin{center}
\includegraphics[width=.5\textwidth]{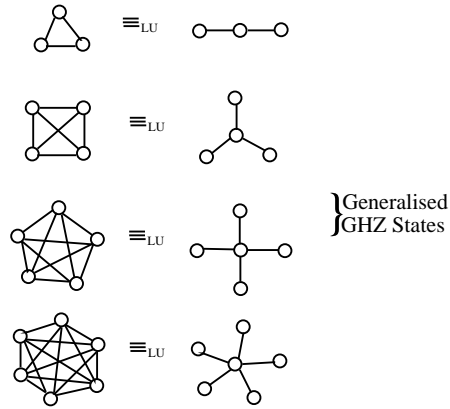}
\caption[]{Fully-Connected $\leftrightarrow$ GHZ Equivalence \label{eps11}}
\end{center}
\end{figure}

\begin{figure}
$$ \begin{array}{ccc}
\begin{tiny} s({\bf{x}}) \end{tiny} & \Leftarrow H(0)H(1) \Rightarrow &
\begin{tiny} s'({\bf{x}})
\end{tiny} \\
\includegraphics[width=.2\textwidth]{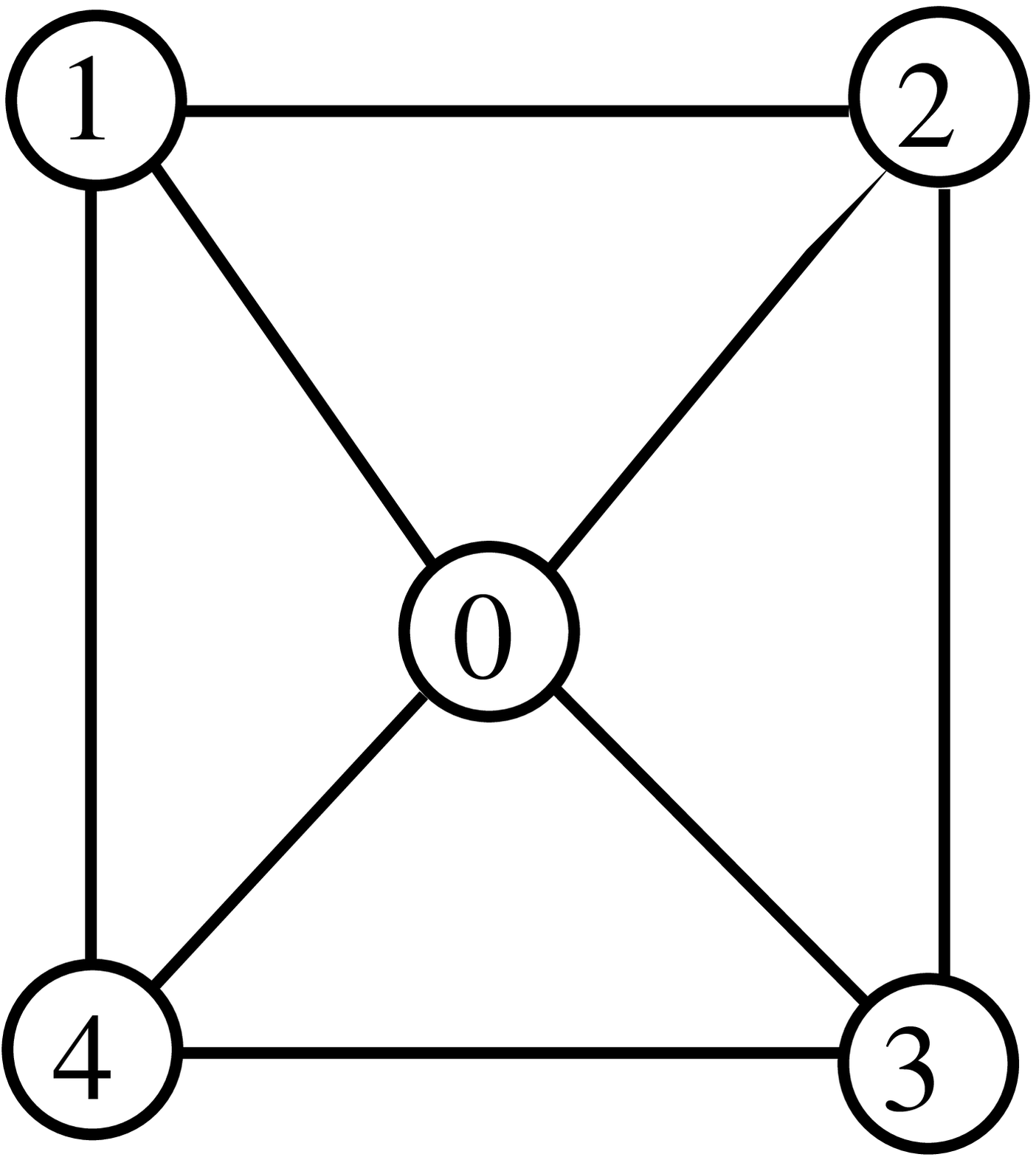} &  &
\includegraphics[width=.2\textwidth]{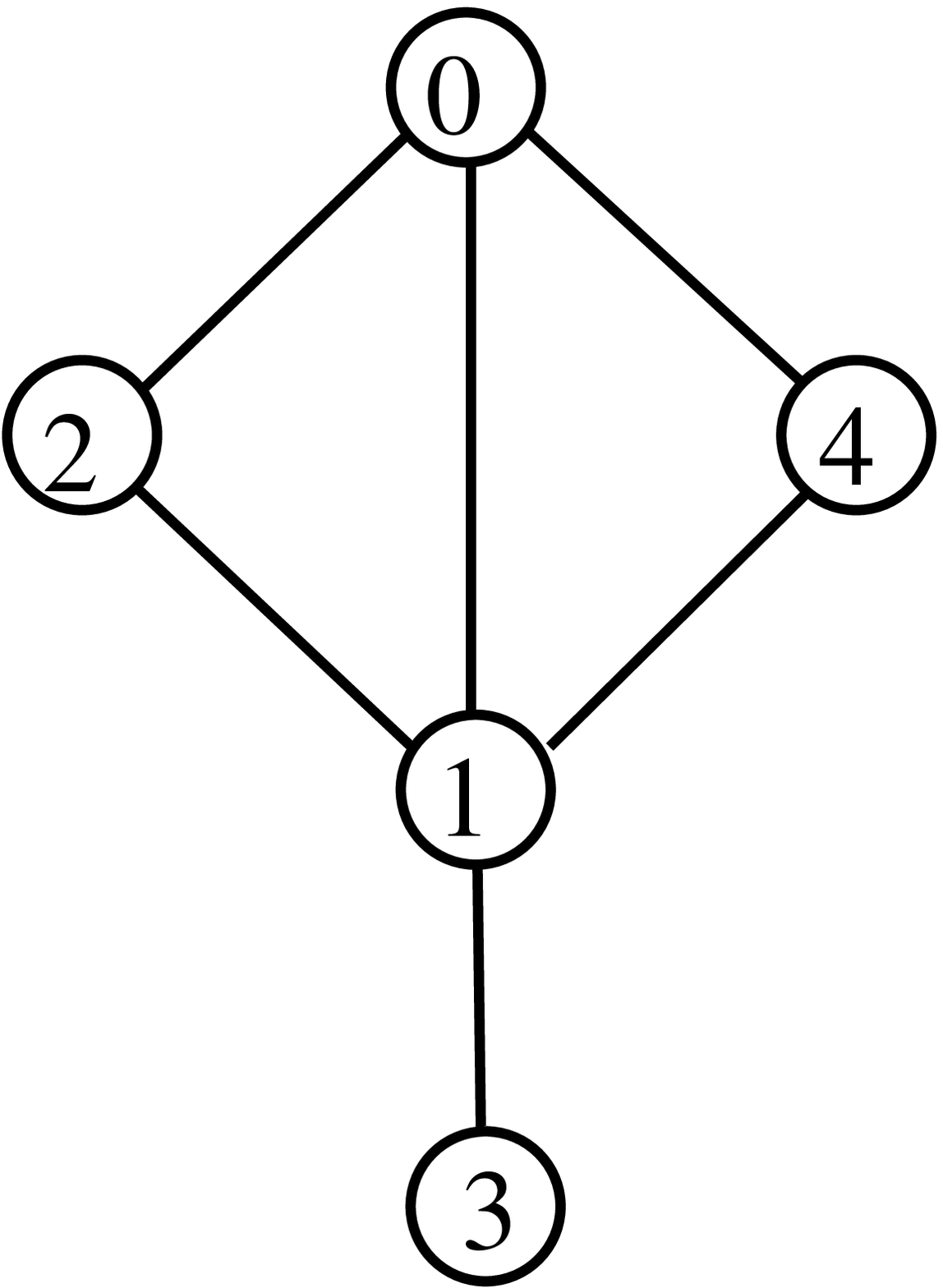} \end{array} $$
\caption[]{LU-Equivalent Non-Bipartite Quadratic Bipolar States \label{eps12}}
\end{figure}

\begin{figure}
\begin{center}
\includegraphics[width=.3\textwidth]{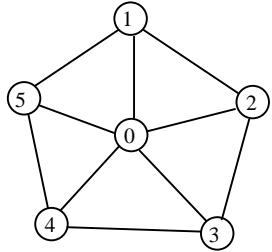}
\caption[]{Quadratic State with Conjectured PAR$_l = 4.486$ \label{eps13}}
\end{center}
\end{figure}

This paper has focussed on states from ${\bf{\ell_p}}$.
In this section we take a preliminary look at quadratic bipolar states which do not
have bipartite form. A 'Fully-Connected Quadratic' is a bipolar quantum state of
the form,
$$ s({\bf{x}}) = (-1)^{\sum_{j = 0}^{n-2} \sum_{i = j+1}^{n-1} x_jx_i} $$
Although these states appear to have a very low PAR of $2.0$ under $HI$ multispectra,
their true PAR$_l$ is $2^{n-1}$.
Fully-Connected States satisfy the following:
\begin{small}
$$ {\bf{s'}} = (10 \ldots 01) = \left ( \begin{array}{cc}
w^7 & 0 \\
0 & w
\end{array} \right )_f \prod_{j=0}^{n-1} {\bf{NH}}(j)\vec{s} $$
\end{small}
where $w = e^{\frac{2\pi i}{8}}$, ${\bf{NH}} = $ \begin{tiny} $\left ( \begin{array}{cc}
1 & i \\
1 & -i
\end{array} \right )$ \end{tiny} (the NegaHadamard Transform \cite{Par:Gol}), $i^2 = -1$,
and ${\bf{s'}}$ is the binary indicator
for the $[n,1,n]$ Generalised GHZ state. \begin{tiny} $\left ( \begin{array}{cc}
w^7 & 0 \\
0 & w
\end{array} \right )_f$ \end{tiny} acts on qubit $f$, where $f$ is chosen arbitrarily
from $\{0,1,\ldots,n-1\}$. So Fully-Connected quadratic
bipolar states are LU equivalent to ${\bf{\ell_p}}$. Moreover, as
$k = 1$ for a Generalised GHZ state, only {\bf{one}} measurement is necessary
(of any qubit in the ${\bf{NH}}$-basis) to completely disentangle the state. We illustrate the
equivalence in Fig \ref{eps11}, where the rh-side shows the bipolar form of the GHZ
state.
\footnote{Fully-Connected quadratic bipolar states seem to be a generalisation of the
'maximal-connectedness' of \cite{Brie:Ent} (see also \cite{Bru:Ent})}
Next we present a quadratic bipolar state which is not LU equivalent to a code
with binary probabilities.  Let
$s({\bf{x}}) = (-1)^{x_0x_1 + x_0x_2 + x_0x_3 + x_0x_4 + x_1x_2 + x_1x_4 + x_2x_3 +
x_3x_4}$. By computation, PAR$_l(\vec{s}) = 8.0$ (e.g. under the WHT), suggesting that
only $\log_2(\frac{32}{8}) = 2$ measurements are necessary to completely disentangle
the state, for instance, by measuring  qubits $0$ and $1$ in the $H$-basis.
The lh-side of Fig \ref{eps12} is $s({\bf{x}})$, and the
rh-side is $s'({\bf{x}}) = (-1)^{x_0x_1 + x_0x_2 + x_0x_4 + x_1x_2 + x_1x_3 + x_1x_4}$,
obtained from $s({\bf{x}})$ by application of $H(0)H(1)$.
Finally, our computations suggest that the quadratic bipolar state described by Fig
\ref{eps13} (or any of its 72 qubit permutations)
has PAR$_l = 4.486$, although at 6 qubits we are working at our computational limit). This
PAR$_l$ is very low - the lowest possible PAR$_l$ for a 6-qubit state from ${\bf{\ell_p}}$
is $2^3 = 8$. No other $6$-qubit bipolar quadratics have PAR$_l$ this low.
No more than 4
measurements are required to completely destroy entanglement in the state. For
instance, by measuring qubits 0,1,3,5 in the bipolar basis.
\section{Measurement-Driven Computation}
\label{sec10}
\begin{figure}
\begin{center}
\includegraphics[width=.2\textwidth]{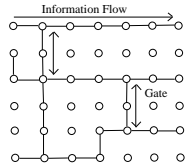}
\caption[]{Quadratically-Entangled Array \label{eps14}}
\end{center}
\end{figure}

\begin{figure}
\begin{center}
\includegraphics[width=.3\textwidth]{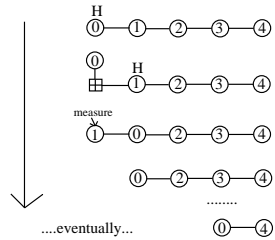}
\caption[]{Quantum State Preparation \label{eps15}}
\end{center}
\end{figure}

Recent research \cite{Brie:Ent,Raus:QC} proposed
entangled arrays of particles to perform Quantum Computation as in Fig \ref{eps14},
using the entangling primitive, \newline
\begin{small}
$ \frac{1}{2^{N/2}} \bigotimes_{a=0}^{N-1} (|0>_a\sigma_z^{(a+1)} + |1>_a) $
\end{small},
where $\sigma_z = $ \begin{tiny} $\left ( \begin{array}{cc} 1 & 0 \\ 0 & -1 \end{array} \right )$. \end{tiny}
The array holds a quadratic bipolar state. Moreover, this
state is from ${\bf{\ell_p}}$ due to the bipartite form of a rectangular array. For such
states, the
Schmidt Measure of Entanglement, discussed in \cite{Raus:QC}, corresponds to the LE
discussed in this paper. The weight hierarchy of this paper further refines the
'Persistency of Entanglement' measure \cite{Brie:Ent} to 'Stubborness of Entanglement'.
Selective measurement drives computation on the array, exploiting
inherent entanglement. We now give an example which was first described in
\cite{Brie:Ent}, but we repeat it here.
It is desired to teleport a quantum state from position
$i$ to position $j$. This can be achieved with certainty if
we possess an entangled pair of qubits, $x_i$ and $x_j$, which are in the
state $x_i + x_j + 1$. By applying $H(i)$ we view this state as $(-1)^{x_ix_j}$.
The problem is therefore to prepare the state $(-1)^{x_ix_j}$. In our
array of qubits there is
always a connective route from qubit $i$ to qubit $j$ for every $i$ and $j$. We measure
all extraneous variables to leave a quadratic line graph between
$x_i$ and $x_j$, ending in $x_i$ and $x_j$. Therefore
$ s({\bf{x}}) = (-1)^{x_ix_0 + x_0x_1 + \ldots + x_{q-2}x_{q-1} + x_{q-1}x_j} $
and we wish to create $(-1)^{x_ix_j}$. Using Theorems \ref{thma2} and \ref{thma5},
the application of $H(q-1)H(j)$ on $\vec{s}$ gives
$ s'({\bf{x}}) = (-1)^{x_ix_0 + x_0x_1 + \ldots + + x_{q-1}x_j + x_{j}x_{q-1}} $.
So $x_j$ and $x_{q-1}$ swap in the expression, (although their
physical position in the array remains unchanged). We now measure $x_{q-1}$ to
disentangle it. (This may produce the extra term $x_j$ if $x_{q-1}$ is measured as $1$, but
we can ignore this 'linear offset'). We repeat the above by applying $H(q-2)$
and $H(j)$, then measuring $x_{q-2}$,...etc, until we are left with $(-1)^{x_ix_j}$. We can
use our 'EPR pair', $(-1)^{x_ix_j}$, to teleport quantum information from position $i$ to
position $j$. This 'state preparation' is shown in Fig \ref{eps15}.
This example is also useful in illustrating the 'danger' of entanglement with the environment.
Consider the bipolar state $s({\bf{x}}) = (-1)^{x_0x_1 + x_1x_e}$, where $x_e$ is an environmental
qubit. It may seem that qubits 0 and 1 are entangled independent of $x_e$ but, as above, we can
swap the positions of $x_1$ and $x_e$ by applying $IHH$ to get
$s'({\bf{x}}) = (-1)^{x_0x_e + x_ex_1 = x_e(x_0 + x_1)}$. Thus ${\bf{\rho_e}}$, the mixed state
of qubits 0 and 1, is separable,
and there is no two-qubit entanglement between qubits 0 and 1.
\section{Conclusion}
\label{sec11}
This paper has approached Quantum Entanglement from a Coding and Sequence
Theory perspective. We argue that optimum and near-optimum binary linear codes are
quantum states with high entanglement. We show how to interpret
coding and sequence parameters in a quantum context. In particular, we have identified
sequence PAR, code dimension, weight hierarchy, nonlinear order, and correlation immunity,
as having useful meanings in a
quantum context. Most importantly, we have shown that, for quantum states which are
equivalent under local unitary transform to indicators for binary linear
error-correcting codes, the partial entanglement measures of Linear Entanglement
and Stubborness of Entanglement can be computed by only examining the multispectra
arising from tensor products of $2 \times 2$ Walsh-Hadamard and Identity matrices.
This allows us to consider entanglement of multiparticle states
over many particles, in the same way that coding theorists consider properties of
large blocklength codes. One implication of this is that current research into
very long blocklength codes, using codes constructed on graphs, could be applied to the
study of Quantum Entanglement \cite{Par:QFG}, and vice versa. Local Unitary
equivalence of quantum states allows us to look at classical coding and sequence design
problems from new angles (literally) and, in future papers, we hope to use the idea of
mixed-state entanglement to study the design and use of codes in certain channels.
We also expect that Local Unitary equivalence will give us a new way of looking
at cryptographic properties of certain sequences.
This paper has concentrated on bipolar quadratic entanglement, in particular
entanglement that can be described using binary linear block codes.
Future work will consider higher-degree and higher-alphabet states, and we
will need to search outside the $HI$ multispectra to find the desired
entanglement parameters.
\section{Appendix: $HI$ Multispectra and the Set, ${\bf{\ell_p}}$}
\label{App}
\subsection{The Action of $H(i)$}
\begin{lem}
\label{lem1}
$ A(-1)^B \Leftrightarrow A(-1)^{AB}$, where $A$ and $B$ are both binary functions.
\end{lem}
\begin{lem}
\begin{small}
$\left ( \prod_{i = 0}^{F-1} a_i \right ) + \left ( \prod_{i = 0}^{F-1} (a_i + 1) \right )
 = \prod_{
\begin{tiny} \begin{array}{c} i = 0 \\ i \ne j \end{array} \end{tiny}}^{F-1}
(a_j + a_i + 1) $ \end{small}, for some arbitrary $j \in \{0,1,\ldots,F-1\}$.
\label{lem3}
\end{lem}
\begin{df}
We use the expression $x_k \in g({\bf{x}})$
($x_k \not \in g({\bf{x}})$) to
indicate that $g({\bf{x}})$ is (is not) a function of $x_k$.
\label{dfa11}
\end{df}
\begin{df}
For a certain $x_i$, let ${\bf{R}}$ be a subset of the integers such
that $x_i \not \in h({\bf{x}})_k$, $\forall k \in {\bf{R}}$. Similarly,
let ${\bf{V}}$ be a subset of the integers such
that $x_i \in h({\bf{x}})_k$, $\forall k \in {\bf{V}}$. Then
$ m({\bf{x}}) = r({\bf{x}})v({\bf{x}}) $,
where $r({\bf{x}}) = \prod_{k \in {\bf{R}}} h({\bf{x}})_k$ and
$v({\bf{x}}) = \prod_{k \in {\bf{V}}} h({\bf{x}})_k$, and where
$r({\bf{x}}) = 1$ if ${\bf{R}} = \emptyset$, and
$v({\bf{x}}) = 1$ if ${\bf{V}} = \emptyset$.
\label{dfa12}
\end{df}
Let $g({\bf{x}})_{| x_i = t}$ mean the polynomial $g({\bf{x}})$
evaluated at $x_i = t$. (In the quantum context this may be considered as
the resultant state of $g$ after a measurement of qubit $i$ in the
$x_i$ basis which resulted in an observation of $t$ for qubit $i$).
\begin{df} $ m_0 = m({\bf{x}})_{| x_i = 0}, \hspace{5mm} \mbox{ }
m_1 = m({\bf{x}})_{| x_i = 1} $,
where $i \in \{0,1,\ldots,n-1\}$.
We similarly define $v_0,v_1,p_0,p_1$ as being the
evaluations of $v({\bf{x}})$, $p({\bf{x}})$,
with $x_i$ fixed to $0$ and $1$, respectively.
\label{dfa1}
\end{df}
\begin{thm}
Let $s = m(-1)^p$. Then the action of $H(i)$ on $s$ is $s'$, where,
\begin{small}
$$ \begin{array}{c}
s' = m'(-1)^{p'} = r((v_0 + v_1) \oplus 2(v_0v_1(p_0 + p_1 + x_i + 1))) \\
\times (-1)^{(v_0v_1((p_0 + 1)(p_0 + p_1 + x_i) + p_0) + v_0p_0 + v_1(p_1 + x_i))}
\end{array} $$ \end{small}
where $\oplus$ indicates conventional, non-modular addition.
\label{thma1}
\end{thm}
\begin{proof}
$ s = m(-1)^{p} = (1 + x_i)m_0(-1)^{p_0} + x_im_1(-1)^{p_1} $,
where $m_0,m_1,p_0,p_1$ are the result of fixing $x_i$. Applying $H(i)$ gives,
\begin{equation} \begin{array}{c}
s' = (1 + x_i)(m_0(-1)^{p_0} \oplus m_1(-1)^{p_1}) +
x_i(m_0(-1)^{p_0} \ominus m_1(-1)^{p_1}) \\
 = (1 + x_i)(m_0(p_0 + 1) \oplus m_1(p_1 + 1) \ominus m_0p_0 \ominus m_1p_1) \\
 + x_i(m_0(p_0 + 1) \ominus m_1(p_1 + 1) \ominus m_0p_0 \oplus m_1p_1)
\end{array} \label{eqa1} \end{equation}
where $\oplus$ and $\ominus$ indicate non-modular addition and subtraction, respectively.
We now use the identity,
\begin{equation} A_0 \oplus A_1 \ominus B_0 \ominus B_1 =
((A_0 + A_1 + B_0 + B_1) \oplus 2(A_0A_1 + B_0B_1))(-1)^{B_0B_1 + B_0 + B_1} \label{eqa2} \end{equation}
where $A_0,A_1,B_0,B_1 \in \{0,1\}$. Applying (\ref{eqa2}) to (\ref{eqa1}) gives,
\begin{equation} \begin{array}{c}
s' = (1 + x_i)((m_0 + m_1) \oplus 2(m_0m_1(p_0 + p_1 + 1))(-1)^{m_0m_1p_0p_1 + m_0p_0 + m_1p_1} \\
+ x_i((m_0 + m_1) \oplus (m_0m_1(p_0 + p_1))(-1)^{m_0m_1p_0(p_1+1) + m_0p_0 + m_1p_1 + m_1}
\end{array} \label{eqa3} \end{equation}
Applying Lemma \ref{lem1} to (\ref{eqa3}) enables us to factor out $r$ from the exponent of
$(-1)$, and more generally reduce the exponent. Finally, we
obtain Theorem \ref{thma1} by observing that $p_0p_1 = p_0(p_0 + p_1 + 1)$.
 \hspace{5mm} \mbox{ }\rule{2mm}{3mm} \end{proof}
We simplify Theorem \ref{thma1} somewhat, as
we are interested in binary APFs (i.e. where $m({\bf{x}})$ and $p({\bf{x}})$ are both binary).
\begin{cor}
$s'({\bf{x}})$ of Theorem \ref{thma1} is a binary APF iff either
$v_0 + v_1 = 0$, or $v_0v_1 = 0$.
(Note that $(p_0 + p_1 + x_i + 1)$ can never be zero).
\label{cor1}
\end{cor}
Consider the case where $v_0 + v_1 = 0$.
\begin{thm}
If $v_0 + v_1 = 0$ then $x_i \not \in m({\bf{x}})$, and vice versa, and
Theorem \ref{thma1} reduces to,
$$ s' = m(c + x_i + 1)(-1)^{p_0} $$
where $c = p_0 + p_1$ is the sum of 'connection' terms, connected by multiplication
to $x_i$ in $p({\bf{x}})$.
\label{thma2}
\end{thm}
\begin{proof}
$v_0 + v_1 = 0$ iff $v_0 = v_1 = 1$. Theorem \ref{thma2} follows by
Lemma \ref{lem1}.
 \hspace{5mm} \mbox{ }\rule{2mm}{3mm} \end{proof}
\begin{small} {\bf{Example: }} $H(2)$ acting on
$s({\bf{x}}) = (x_0 + x_3)(x_0 + x_1 + 1)(-1)^{x_0x_1x_3 + x_2x_3 + 1}$ gives
$s'({\bf{x}}) = (x_0 + x_3)(x_0 + x_1 + 1)(-1)^{x_0x_1x_3 + 1}$. \end{small} \newline
We now consider $v_0v_1 = 0$.
Let $h_{0,k} = h({\bf{x}})_{k_{| x_i = 0}}$ and
$h_{1,k} = h({\bf{x}})_{k_{| x_i = 1}}$.
\begin{thm}
$v_0v_1 = 0$ iff $m({\bf{x}})$ contains at least
one term $h({\bf{x}})_k$ which has a linear dependence on $x_i$.
In this case, Theorem \ref{thma1} reduces to,
$$ s' = r(v_0 + v_1)(-1)^{p_0 + v_1(c + x_i)} $$
where $c = p_0 + p_1$.
\label{thma21}
\end{thm}
\begin{proof}
If $h({\bf{x}})_k$ is linear in $x_i$ for some
$k$, then $h_{1,k} = h_{0,k} + 1$,
and $h_{1,k}h_{0,k} = 0$.
$h_{0,k}$ is a factor
of $v_0$, and $h_{1,k}$ is a factor of $v_1$, therefore $v_0v_1 = 0$.
Theorem \ref{thma1} initially reduces to
$ s' = r(v_0 + v_1)(-1)^{v_0p_0 +v_1(p_1 + x_i)} $. Therefore
$ s' = r(v_0 + v_1)(-1)^{(v_0 + v_1)p_0 + v_1(c + x_i)}$.
Applying Lemma \ref{lem1} gives Theorem \ref{thma21}.
 \hspace{5mm} \mbox{ }\rule{2mm}{3mm} \end{proof}
\begin{small}
{\bf{Example: }} $H(2)$ acting on
$s({\bf{x}}) = (x_2 + x_3 + 1)(x_0x_1 + 1)(x_0x_2)(-1)^{x_0x_3 + x_1x_2x_3}$ gives
$s'({\bf{x}}) = (x_0x_1 + 1)(x_0x_3)(-1)^{x_0x_1x_3 + x_0x_2x_3 + x_0x_3}$.
\end{small}
\begin{thm}
Let $m({\bf{x}})$ be chosen so that $x_i \in m({\bf{x}})$ and so that, if $x_i \in h({\bf{x}})_k$, then $h({\bf{x}})_k$
is linear in $x_i$. Then Theorem \ref{thma1} reduces to a special case of Theorem
\ref{thma21}, where,
$$ s' = r(v_0 + v_1)(-1)^{p_0 + h_{1,z}(c + x_i)}, \hspace{5mm} \mbox{ }z \in {\bf{V}} $$
where $c = p_0 + p_1$, $z$ is chosen arbitrarily from ${\bf{V}}$, and, \newline
\begin{small}
$ (v_0 + v_1) = \prod_{\begin{tiny} \begin{array}{l} k \in {\bf{V}} \\ k \ne j \end{array} \end{tiny}
} (h_{0,j} + h_{0,k} + 1), \hspace{5mm} \mbox{ }j \in {\bf{V}} $ \end{small},
with $j$ chosen arbitrarily from ${\bf{V}}$. We can make $j = z$ if we want to.
\label{thma5}
\end{thm}
\begin{proof}
\begin{small}
$ v_0 + v_1 = \left ( \prod_{k \in {\bf{V}}} h_{0,k} \right ) + \left ( \prod_{k \in {\bf{V}}} h_{1,k} \right
) $ \end{small}.
But, for $h_k({\bf{x}})$ linear in $x_i$, then $h_{1,k} = h_{0,k} + 1$. The expression
for $v_0 + v_1$ in Theorem
\ref{thma5} follows by Lemma \ref{lem3}. $v_0 + v_1$ is only $1$ when $h_{1,k} = 1$ $\forall k$, or
$h_{1,k} = 0$ $\forall k$. Moreover, by definition, $v_1 = 1$ only when $h_{1,k} = 1$ $\forall
k$. We can therefore replace $v_1$ in the exponent of Theorem \ref{thma5} by any function
which is $1$ for $h_{1,k} = 1$ $\forall k$, and $0$ for $h_{1,k} = 0$ $\forall k$.
We choose to replace $v_1$ with $h_{1,z}$, as described in Theorem
\ref{thma5}, where $v$ is arbitrarily chosen from ${\bf{V}}$.
 \hspace{5mm} \mbox{ }\rule{2mm}{3mm} \end{proof}
\begin{small}
{\bf{Example: }} $H(2)$ acting on
$s({\bf{x}}) = (x_1 + x_3)(x_0 + x_1 + x_2 + 1)(-1)^{x_0x_1x_2 + x_1x_3 + 1}$ gives
$s'({\bf{x}}) = (x_0 + x_2 + x_3)(-1)^{x_0x_2x_3 + x_0x_2 + x_1x_3 + x_1 + 1}$, where
$h_{1,z}$ was chosen as $(x_3 + 1)$.
\end{small}

\begin{proof}
({\bf{Theorem \ref{thma4}}})
Let $s({\bf{x}}) = m({\bf{x}})(-1)^{p({\bf{x}})}$ be a binary spectra APF.
If $x_i \not \in m({\bf{x}})$, then Theorem \ref{thma2} tells us that
$\vec{s'}$ is a binary APF where
$m({\bf{x}})$ becomes $m'({\bf{x}}) = m({\bf{x}})(c + x_i + 1)$. Also $\deg(c) \le 1$,
so $m'({\bf{x}})$ has a linear product decomposition and, from Theorem \ref{thma2}, $p({\bf{x}})$ becomes
$p_0$, where $\deg(p_0) \le 2$. Therefore, if
$x_i \not \in m({\bf{x}})$ then $\vec{s'}$ has a binary spectra APF.
Secondly, if $x_i \in m({\bf{x}})$, then Theorem \ref{thma21} tells us that $v_0v_1 = 0$,
as at least one $h({\bf{x}})_k$ in $m({\bf{x}})$ is linear in $x_i$. Theorem \ref{thma5} tells us
that, because all $h({\bf{x}})_k$ are of degree one, then $\vec{s'}$ is a binary
APF. This is because
$m'$ is still a product of degree one terms (i.e. $v_0 + v_1$ is a product of degree one terms).
As both $h_{1,z}$ and $c + x_i$ are of degree one, then
$\deg(p_0 + h_{1,z}(c + x_i)) \le 2$. So, once again, $\vec{s'}$ has a binary spectra APF.
 \hspace{5mm} \mbox{ }\rule{2mm}{3mm} \end{proof}
{\bf{Example: }} $H(0)$ acting on
$ s({\bf{x}}) = (x_0 + x_1 + x_2 + x_3 + 1)(x_0 + x_1 + x_6)(x_0 + x_4 + 1)(-1)^{x_0x_5}$ gives
$ s'({\bf{x}}) = (x_2 + x_3 + x_6 + 1)(x_1 + x_2 + x_3 + x_4 + 1)(-1)^{(x_1 + x_2 + x_3)(x_0 + x_5)} $,
where we have chosen $h_{0,j} = (x_1 + x_2 + x_3 + 1)$ and
$h_{1,z} = (x_1 + x_2 + x_3)$.
We could alternatively choose, say, $h_{0,j} = (x_1 + x_2 + x_3 + 1)$ and
$h_{1,z} = (x_4)$, in which case
$s'({\bf{x}}) = (x_2 + x_3 + x_6 + 1)(x_1 + x_2 + x_3 + x_4 + 1)(-1)^{(x_4)(x_0 + x_5)} $.
These two expressions for $\vec{s'}$ are equivalent.

\begin{proof}
({\bf{Theorem \ref{thma6}}}) Let $\vec{s} \in {\bf{\ell_p}}$.
We wish to reduce the exponent of $(-1)$ of $\vec{s}$ to zero by judicial applications
of multiple $H(i)$. From Theorem \ref{thma2}, the action of $H(i)$ on $\vec{s}$
gives $s'({\bf{x}})$, with
exactly one linear term, $m'({\bf{x}})$, namely $(c + x_i + 1)$. Let $x_j \in c$.
From Theorem \ref{thma5} the subsequent action of $H(j)$ on $s'({\bf{x}})$ puts $x_j$
back into the exponent of $(-1)$, where it is 'reconnected' via multiplication to the
term $h_{1,z}$. But in this case $h_{1,z}$ must be $(c + x_i + 1)_{| x_j = 1}$,
so the exponent must include the quadratic term $x_ix_j$. We have already used the
actions $H(i)$ and $H(j)$, and no subsequent $H(k)$ actions, $k \ne j$, can remove
this $x_ix_j$ term from the exponent. Thus, if $x_i$ and $x_j$ are connected via
multiplication in $p({\bf{x}})$, then $i$ and $j$ cannot both be part of the set ${\bf{T_C}}$
(or ${\bf{T_{C^{\perp}}}}$).
Secondly, let the term $x_ix_0 + x_0x_1 + \ldots + x_{q-1}x_j$ occur in $p({\bf{x}})$
where, without loss of generality, we assume $i,j \not \in \{0,1,\ldots,q-1\}$. Let us
apply $H(i)$ and $H(j)$ but not apply $H(0)$, $H(1)$, \ldots, $H(q-1)$.
Then, from Theorem \ref{thma2}, $x_0x_1 + \ldots + x_{q_2}x_{q-1}$ remains in the exponent of $(-1)$.
Elimination of these terms from the exponent is ensured if $q \le 1$.
The above two restrictions together imply that for each term of the form $x_ix_k + x_kx_j$
in $p({\bf{x}})$, either $H(i)$ and $H(j)$ must be applied or $H(k)$ must be applied to
ensure a final $s'({\bf{x}}) = m'({\bf{x}})$. This automatically implies a bipartite
splitting, where $x_i,x_j \in {\bf{T_C}}$ and $x_k \in {\bf{T_{C^{\perp}}}}$.
From Theorem \ref{thma2} the application of $\prod_{i \in {\bf{T_C}}} H(i)$ to
$\vec{s}$
must give a final $p_0$ equal to zero.
To understand why these bipartite transformations reach all linear ECCs,
we note that $s'({\bf{x}}) = m'({\bf{x}})$ must satisfy
$ s'({\bf{x}}) = \prod_{i \in {\bf{T_C}}} (c_i + x_i + 1) $,
where $c_i$ is the linear sum of the variables connected by multiplication to $x_i$ in
$p({\bf{x}})$. We could write each term $(c_i + x_i + 1)$ as a row of a parity check
matrix, and in this way recover a conventional description for a linear
ECC: Each $(c_i + x_i + 1)$ must be $1$ for $s'({\bf{x}})$ to be $1$. This translates to
each parity check row $(c_i + x_i)$ being equal to zero for a valid codeword.
 \hspace{5mm} \mbox{ }\rule{2mm}{3mm} \end{proof}
\section{Appendix: PAR$_l$ and Linear Entanglement (LE)}
\label{App1}
\begin{proof} ({\bf{Theorem \ref{thmX2}}}) Let $H_{e_i}$, $U_{e_i}$, and $I_{e_i}$ be the PARs of the resultant
spectrum, $\vec{s'}$, after application of $H(e_i)$, $U(e_i)$, or $I(e_i)$, respectively, on
$\vec{s}$, where $U(e_i)$ is any $2 \times 2$ unitary
matrix acting on qubit $x_{e_i}$, and $I(e_i)$ is the $2 \times 2$ identity matrix acting
on $x_{e_i}$.
The maximum possible PAR scaling of $\vec{s'}$ relative to $\vec{s}$
after application of one $2 \times 2$ unitary matrix is by a factor of 2, either up or
down.
But from Lemma \ref{lemZ5}, when $\vec{s} \in {\bf{\ell_p}}$,
the application of $H_{e_i}$ on $\vec{s'}$
scales PAR by a factor of 2, upwards or downwards. We therefore conclude that,
\begin{equation} \begin{array}{c}
\mbox{either } U_{e_i} \ge H_{e_i} \hspace{3mm} \mbox{if } H_{e_i} = \frac{1}{2}I_{e_i}
\\ \mbox{or } U_{e_i} \le H_{e_i} \hspace{3mm} \mbox{if } H_{e_i} = 2I_{e_i}
\end{array} \label{eqXX1} \end{equation}
Moreover, \newline
\begin{equation} \frac{1}{2}I_{e_i} \le U_{e_i} \le 2I_{e_i} \label{eqXX3} \end{equation}
Combining (\ref{eqXX1}) and (\ref{eqXX3}) gives,
\begin{equation} \begin{array}{l}
\mbox{if } H_{e_i} = \frac{1}{2}I_{e_i} \mbox{ then } H_{e_i} \le U_{e_i} \le 2I_{e_i} \\
\mbox{if } H_{e_i} = 2I_{e_i} \mbox{ then } \frac{1}{2}I_{e_i} \le U_{e_i} \le H_{e_i}
\end{array}
\label{eqXX4} \end{equation}
 Let us write
 $$ U(e_i) = \left [ U(e_i)H(e_i) \right ] H(e_i)  = V(e_i)H(e_i)$$
 for some $2 \times 2$ unitary matrix, $V(e_i)$. Then,
\begin{equation} \frac{1}{2}H_{e_i} \le U_{e_i} \le 2H_{e_i} \label{eqXX2} \end{equation}
Combining (\ref{eqXX4}) and (\ref{eqXX2}) gives,
\begin{equation} \begin{array}{l}
\mbox{if } H_{e_i} = \frac{1}{2}I_{e_i} \mbox{ then } H_{e_i} \le U_{e_i} \le I_{e_i} \\
\mbox{if } H_{e_i} = 2I_{e_i} \mbox{ then } I_{e_i} \le U_{e_i} \le H_{e_i}
\end{array}
\label{eqX1} \end{equation}
Let any product of the $H_{e_i}$, $U_{e_j}$, and $I_{e_k}$ define the resultant PAR
after the application of the same tensor product of $H(e_i)$, $U(e_j)$, and $I(e_k)$.
Then we can write,
\begin{equation} \mbox{either } H_{e_j}U_{e_i} \le U_{e_j}U_{e_i} \le I_{e_j}U_{e_i},
\hspace{2mm} \mbox{or } I_{e_j}U_{e_i} \le U_{e_j}U_{e_i} \le H_{e_j}U_{e_i}, \hspace{2mm} i \ne j \label{eqX2} \end{equation}
Combining (\ref{eqX1}) and (\ref{eqX2}) guarantees that at least one of the four following
conditions is satisfied:
$$ \begin{array}{ll}
\mbox{either } & H_{e_j}H_{e_i} \le U_{e_j}U_{e_i} \le I_{e_j}I_{e_i} \\
\mbox{or } & H_{e_j}I_{e_i} \le U_{e_j}U_{e_i} \le I_{e_j}H_{e_i} \\
\mbox{or } & I_{e_j}H_{e_i} \le U_{e_j}U_{e_i} \le H_{e_j}I_{e_i} \\
\mbox{or } & I_{e_j}I_{e_i} \le U_{e_j}U_{e_i} \le H_{e_j}H_{e_i}
\end{array}$$
Continuing in this fashion, it follows that,
$$ \prod_{i \in {\bf{T \setminus t_*}}} H_{e_i}\prod_{j \in {\bf{t_*}}} I_{e_i}
\le \prod_{i \in {\bf{T}}} U_{e_i} \le
 \prod_{i \in {\bf{T \setminus t_*}}} I_{e_i}\prod_{j \in {\bf{t_*}}} H_{e_i} $$
for any integer set ${\bf{T}}$ where ${\bf{t_*}} \subset {\bf{T}}$.
In words, for $\vec{s} \in {\bf{\ell_p}}$,
the PAR of any LU transform of $\vec{s}$ is always upper and lower bounded
by two points in the $HI$ multispectra.  \hspace{5mm} \mbox{ }\rule{2mm}{3mm} \end{proof}

\section{Appendix: Weight Hierarchy and Stubborness of Entanglement (SE)}
\label{App2}
\begin{proof} ({\bf{Theorem \ref{thmQ}}})
Decompose $\vec{s_c}$ as,
$$ s_c({\bf{x}}) = f_0(x_t = 0,g_0({\bf{x}} \backslash x_t)) +
f_1(x_t = 1,g_1({\bf{x}} \backslash x_t)) $$
for boolean functions $f_i$ and $g_i$. From Lemma \ref{lemZ5}, let,
$$ \gamma_t = \frac{\mbox{PAR}(H(t)[\vec{s_c}])
}{\mbox{PAR}(\vec{s_c})} \in \{\frac{1}{2},2\} $$
Then $\gamma_t = 2$ iff $g_0 = g_1$, which is the same as saying that
${\bf{C}}$ has Hamming Distance $d = 1$. Otherwise $\gamma_t = \frac{1}{2}$.
More generally we can find a subset, ${\bf{x_w}}$, of ${\bf{x}}$
such that, if ${\bf{y_w \subset x_w}}$, then
$$
\frac{\mbox{PAR}(\prod_{t \in {\bf{y_w}}}H(t)[\vec{s_c}])
}{\mbox{PAR}(\vec{s_c})} =
\begin{array}{ll} 2^{-|{\bf{y_w}}|}, \hspace{3mm} & {\bf{y_w \ne x_w}} \\
2^{2-|{\bf{y_w}}|} = 2^{2-d_w}, \hspace{3mm} & {\bf{y_w = x_w}} \end{array}
$$
where $d_w$ is any member of the weight distribution of ${\bf{C}}$.
Indeed, for a smallest-size non-empty subset, ${\bf{x_w}}$, we have $d_w = d_1 = d$,
which is the Hamming Distance of ${\bf{C}}$.
Similar arguments hold for the higher weights, $d_2,d_3,\ldots,d_k$ where,
at each stage, a {\underline{smallest}} unused non-empty subset, ${\bf{x_{w_i}}}$ is
added. Let ${\bf{Q}} = {\bf{y_{w_j}}} \cup \bigcup_{i=0}^{j-1} {\bf{x_{w_i}}}$.
After further manipulation we arrive at the Theorem.
\hspace{5mm} \mbox{ }\rule{2mm}{3mm} \end{proof}

\begin{proof} ({\bf{Theorem \ref{thmW}}})
Let $\vec{s_c}$ be the binary indicator for a binary linear ECC.
$\vec{s_c}$ takes one of two forms:
$$ \begin{array}{l}
\mbox{Form } 1. \hspace{3mm} s_c({\bf{x}}) = f_0(x_i = 0,g_0({\bf{x}} \backslash x_i)) +
f_1(x_i = 1,g_1({\bf{x}} \backslash x_i)) \\
\mbox{Form } 2. \hspace{3mm} s_c({\bf{x}}) = f_0(x_i = \gamma,g({\bf{x}} \backslash x_i))
\end{array} $$
where the $f_i,g_i$ are boolean functions, and $\gamma$ is fixed at 0
or 1. After measurement of $x_i$ we obtain
$s'_c({\bf{x}}) = s_c({\bf{x}})_{x_i=\gamma}$ which is,
\begin{itemize}
\item Destructive Measurement - when $\vec{s_c}$ is of Form 1.
\item Redundant Measurement - when $\vec{s_c}$ is of Form 2.
\end{itemize}
The PAR relationships follow straightforwardly by Parseval's Theorem.
Now, consider measuring $s'({\bf{x}}) = H(i)[s_c({\bf{x}})]$. If
$\vec{s_c}$ is of Form 1. then $\vec{s'}$ is of Form 1. or 2.
If $\vec{s_c}$ is of Form 2. then $\vec{s'}$ is of Form 1.
In either case $\vec{s'}$ is still one of the same two
forms and we have the same two measurement scenarios for $\vec{s'}$
as for $\vec{s_c}$.
The proof follows recursively for the whole set, ${\bf{\ell_p}}$. 
\hspace{5mm} \mbox{ }\rule{2mm}{3mm} \end{proof}

\clearpage
\addcontentsline{toc}{section}{Index}
\flushbottom
\printindex

\end{document}